\documentclass[aps,reprint,groupedaddress,showpacs]{revtex4-1}

\usepackage{euscript,amsmath,amssymb,amsfonts,graphicx,color}
\usepackage{epstopdf}

\newcommand{\e}{{\rm e}}
\renewcommand{\d}{{\rm d}}
\newcommand{\pd}{\partial}
\newcommand{\R}{{\mathbb R}}

\newcommand{\x}{{\mathbf x}}
\newcommand{\y}{{\mathbf y}}

\newcommand{\D}{\displaystyle}

\newcommand{\mc}{\mathcal }
\newcommand{\ve}{\varepsilon}

\renewcommand{\v}{{\mathbf v}}

\newcommand{\sign}{{\rm sign}}
\newcommand{\ep}{\epsilon}

\begin{document}

\title{Neural field model of memory-guided search}
\author{Zachary P. Kilpatrick$^{1,2}$ and Daniel B. Poll$^{3,4}$}
\email{zpkilpat@colorado.edu; daniel.poll@northwestern.edu}
\affiliation{1. Department of Applied Mathematics, University of Colorado, Boulder CO 80309 \\ 2. Department of Physiology \& Biophysics, University of Colorado School of Medicine, Aurora CO 80045 \\
3. Department of Mathematics, University of Houston, Houston TX 77204 \\
4. Department of Engineering Sciences \& Applied Mathematics, Northwestern University, Evanston IL 60208}
\date{\today}

\begin{abstract}
Many organisms can remember locations they have previously visited during a search. Visual search experiments have shown exploration is guided away from these locations, reducing the overlap of the search path before finding a hidden target. We develop and analyze a two-layer neural field model that encodes positional information during a search task. A position-encoding layer sustains a bump attractor corresponding to the searching agent's current location, and search is modeled by velocity input that propagates the bump. A memory layer sustains persistent activity bounded by a wave front, whose edges expand in response to excitatory input from the position layer. Search can then be biased in response to remembered locations, influencing velocity inputs to the position layer. Asymptotic techniques are used to reduce the dynamics of our model to a low-dimensional system of equations that track the bump position and front boundary. Performance is compared for different target-finding tasks.
\end{abstract}

\pacs{87.19.lq,87.10.Ed,87.19.lj,87.19.lr}

\maketitle

\section{Introduction}

Most motile organisms rely on their ability to search~\cite{hills15}. The processes of identifying habitats, food sources, mates, and predators makes use of visual search combined with spatial navigation~\cite{mueller08,kanitscheider17}. One guiding principle used to evaluate how organisms implement search is the exploration-exploitation trade-off~\cite{daw06,cohen07}. {\em Exploiting} one's current position to search locally has a low cost, whereas {\em exploring} by moving to another search position has a higher cost but a potentially higher reward~\cite{mehlhorn15}. Many organisms have developed search strategies that attempt to manage this trade-off in a robust way~\cite{simpson04,hills12}. Mathematical models of search can quantify the resources expended and yielded by different strategies, showing how managing the explore-exploit trade-off is key~\cite{charnov06,bartumeus09}.

Memoryless stochastic processes are commonly used to model the dynamics of searching organisms~\cite{benichou11}. Such models prescribe equations for an agent that moves according to pure diffusion~\cite{andrews04}, mixed advection-diffusion~\cite{benichou05,newby10}, or even local diffusion punctuated by large deviations in position~\cite{viswanathan96}. None of these search strategies rely on information about previous locations the agent has visited. However, some studies, particularly studies that focus on visual search, have examined the impact of memory on random search processes~\cite{thornton07}.  There is evidence suggesting organisms tend to guide their gaze away from locations they have already examined~\cite{posner84,klein88}, a mechanism often called {\em inhibition-of-return (IOR)}. While the degree to which IOR facilitates exploration continues to be debated~\cite{horowitz98,wang10,smith11}, recent studies suggest return saccades to previously visited locations are less frequent than they would be for a memoryless search process~\cite{bays12}.

Neural mechanisms underlying memory of previous visual targets are relatively unknown, but fMRI studies in humans have shed light on specific brain areas that might be involved~\cite{mayer04}. One candidate region is the superior colliculus (SC), known to be involved in oculomotor planning~\cite{horwitz99}. Unilateral lesioning of SC disables IOR during visual search performed in the visual hemifield linked to the damaged half of the brain~\cite{sapir99}. Other oculomotor programming areas linked to IOR include the supplementary eye field (SEF) and frontal eye field (FEF)~\cite{lepsien02,ro03}. Detectable activation of these regions suggests models of IOR should represent previous search locations using some form of persistent neural activity. In this way, there is a parallel between inhibitory tagging of previous locations and working memory, which also engages persistent activity for its implementation~\cite{curtis03,oh04}. However, it remains unclear how persistent activity might be initiated and utilized by a neural circuit to successfully implement search guided by IOR. 

\begin{figure*}
\begin{center} \includegraphics[width=16.5cm]{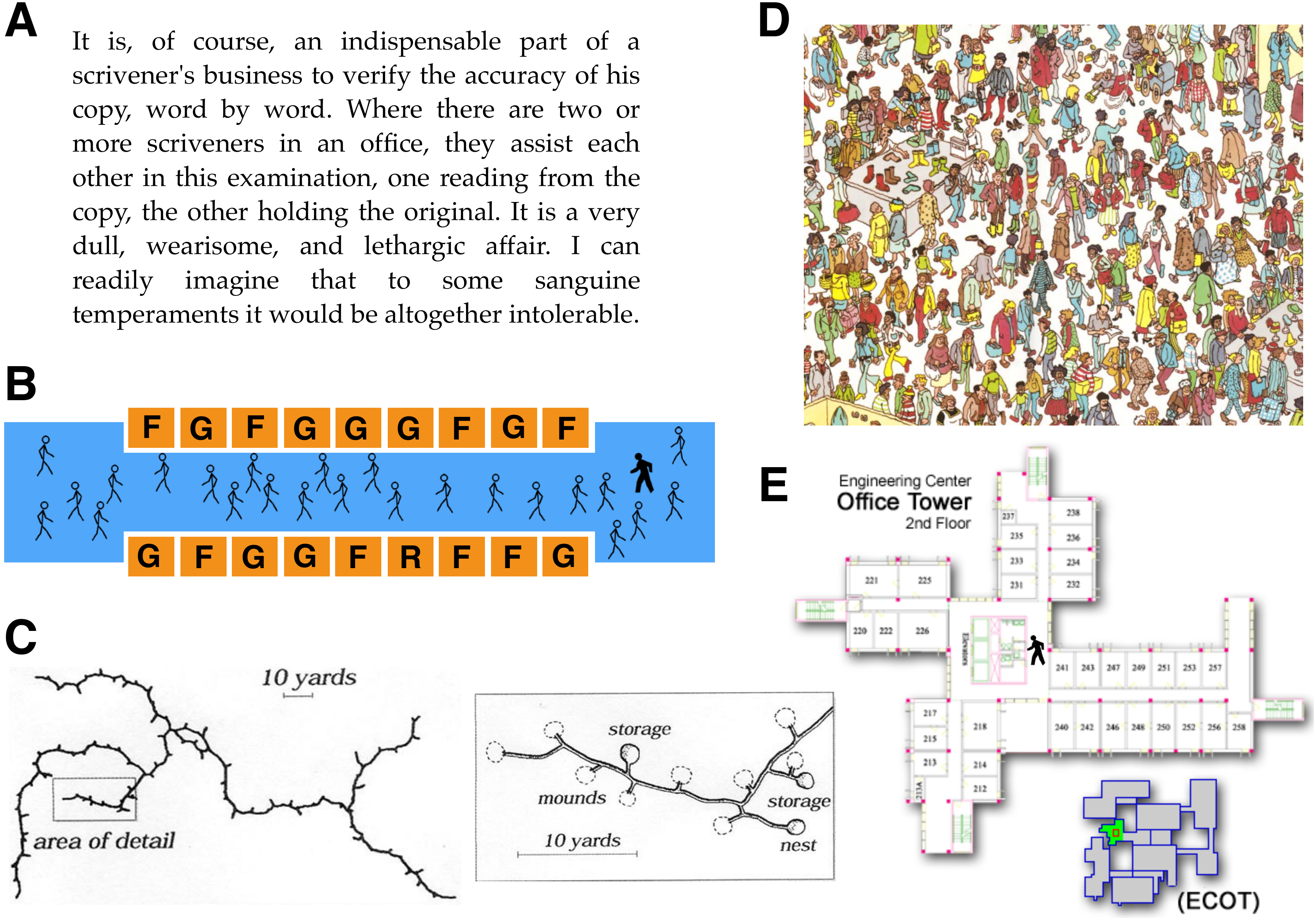} \end{center}
\vspace{-5mm}
\caption{Examples of search tasks. Effectively one-dimensional: {\bf A.} Find the number in the above quote from {\em Bartleby the Scrivener} by Herman Melville (1853)~\cite{melville99}. {\bf B.} Bolded agent must find the the restroom (R) among the food vendors (F) and gates (G) in a narrow airport terminal. {\bf C.} Gopher must find its nest within a narrow network of underground tunnels~\cite{gary17}. Multi-dimensional: {\bf D.} Find Waldo amongst a crowd of non-Waldo visual distractors~\cite{handford87}. {\bf E.} Student looking for their professor's office in the multi-floor (three-dimensional) Engineering Center Office Tower at University of Colorado Boulder~\cite{ecot}.}
\label{fig1_examples}
\end{figure*}

We develop a model of memory-guided search that stores an agent's present search location as well as previously visited search locations. Our model consists of two neural field equations: one neural field layer captures the position of the searcher and the other describes the memory of visited locations. To systematically analyze our neural field model, we focus on one-dimensional search tasks such as scanning lines of text for a specific word (Fig. \ref{fig1_examples}A); searching a long corridor (Fig. \ref{fig1_examples}B); or foraging in a tunnel system (Fig. \ref{fig1_examples}C). We idealize these examples by considering the problem of searching along a one-dimensional segment, or a radial arm maze. After our analysis of the one-dimensional system, we discuss how our model can be extended to two-dimensional search tasks, such as finding an object in a picture (Fig. \ref{fig1_examples}D) or higher dimensions (Fig. \ref{fig1_examples}E).

We proceed by presenting our one-dimensional model of memory of visited spatial locations in Section \ref{model}. Subsequently, we analyze the existence and stability of stationary solutions to our equations in Section \ref{1D}. The position of the searching agent is represented by a stationary pulse (bump) solution in the absence of velocity inputs, and the previously searched region is bounded by two stationary front interfaces. Our analysis provides us with intuition as to how model parameters shape the spatial resolution and robustness of the memory representation. After this, we carry out a low-dimensional reduction of our model, so that we can capture its dynamics by tracking the location of the bump and front interfaces (Section \ref{lowdim}). In fact, this captures the dynamics of the full neural field quite well. Subsequently, we evaluate the performance of memory-guided search, when the velocity inputs are shaped by the previously visited locations stored in memory (Section \ref{perform}). Memory-guided search does not improve the speed of searches along a single segment, but does improve search across multiple connected segments, as in a radial arm maze. Lastly, we discuss extensions to higher dimensions in Section \ref{2D}.





\section{Multilayer neural field model}
\label{model}

\begin{figure*}
\begin{center} \includegraphics[width=16cm]{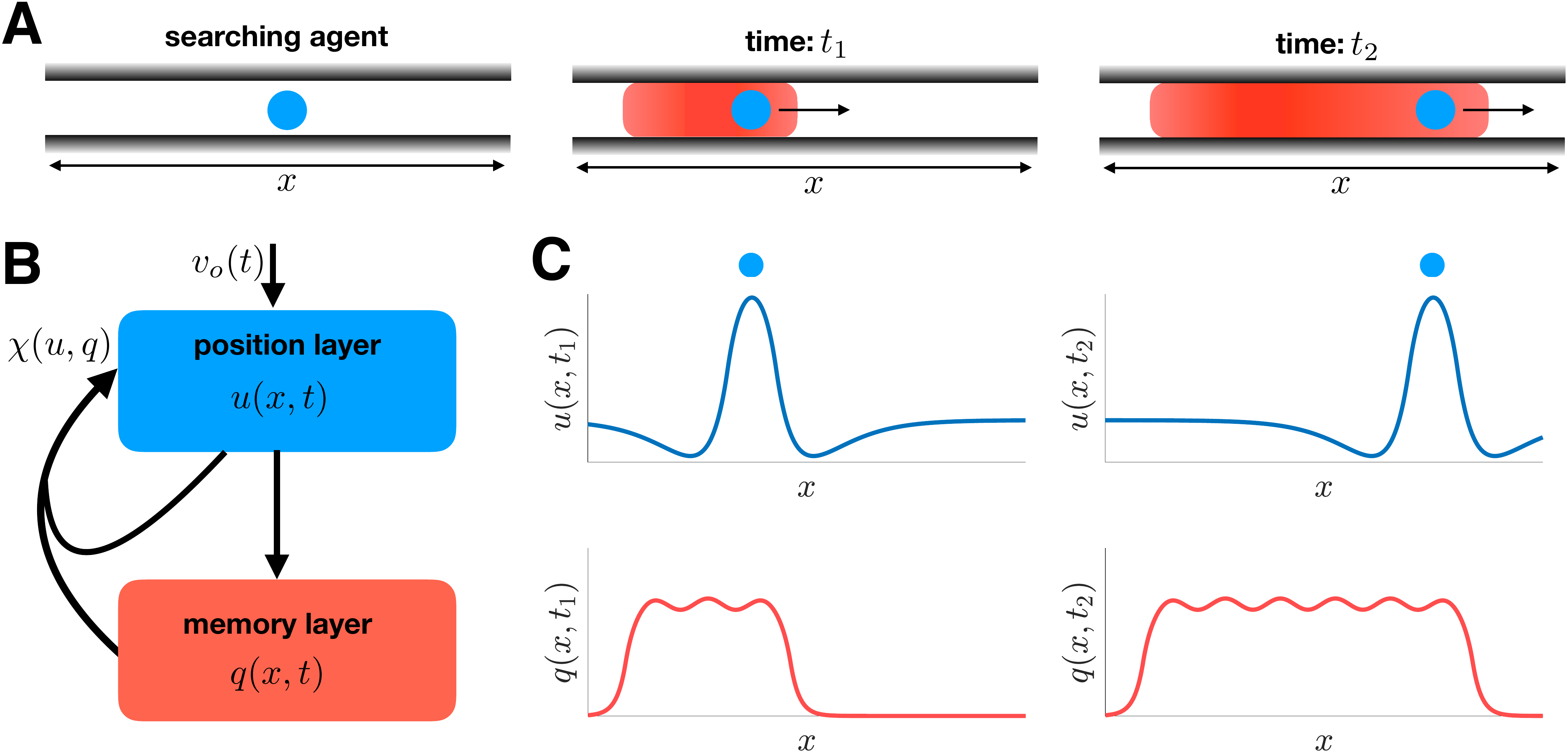} \end{center}
\vspace{-4mm}
\caption{Multi-layer neural field models searching agent's memory. {\bf A.} Searching agent (blue dot) moves along a one-dimensional domain, storing a memory of its previously visited locations (red regions). {\bf B.} Multi-layer neural field model represents current position ($u(x,t)$) and previously visited locations ($q(x,t)$). Both layers shape a closed-loop feedback control to the position layer $\chi (u,q)$, which accounts for previously visited locations. {\bf C.} Peak of bump in position layer ($u(x,t)$) represents agent's current position. Input to memory layer activates neural populations ($q(x,t)$) that represent previously visited locations.}
\label{fig2_model}
\end{figure*}

Our model of memory-guided search assumes uses two layers of stimulus-tuned neurons corresponding to locations of the searching agent. This is mostly motivated by studies of visual search~\cite{wang10}, but there may also be IOR mechanisms that shape search driven by organisms' idiothetic navigation~\cite{hills15}. The first layer of the network encodes the agent's position, and is driven by velocity input, which the network integrates (Section \ref{lowdim}). This layer projects to the second layer, which encodes a memory of locations the agent has visited (Fig. \ref{fig2_model}). Closed-loop control of the velocity input can be implemented by using the memory and position layer to determine the agent's next search location. We discuss this in more detail in Section \ref{perform}. \\
\vspace{-3mm}

\noindent
{\bf Position-encoding layer.} Both the direction of visual gaze~\cite{aksay01} and an animal's idiothetic location~\cite{mcnaughton06} are known to be encoded by position-tuned cells~\cite{burak09,engbert11}. A network ensemble of such neurons would likely rely upon excitatory coupling between cells with similarly-tuned stimulus-preference, and effective inhibition between cells with dissimilar stimulus-preference~\cite{amari77,seung96}. Together with well accepted models of velocity input to networks with positional memory~\cite{zhang96}, this suggests the following neural field model for the layer that supports a bump attractor representing positional memory:
\begin{align}
u_t &= -u +w_u*H(u-\theta_u)- v(t) \left(w_u' \right)*H(u-\theta_u),  \label{pfield}
\end{align}
where $u(x,t)$ is normalized synaptic input in the position-encoding layer at location $x \in (- \infty, \infty)$ at time $t \in [0, \infty)$. Recurrent coupling is described by the integral $w*H(u - \theta_u) = \int_{- \infty}^{\infty} w(x-y) H(u(y,t) - \theta_u) \d y$ with synaptic weight kernel $w(x-y)$, an even-symmetric function of the distance between locations $x$ and $y$, which is locally excitatory ($w>0$: $|x-y|<r$) and distally inhibitory ($w<0$: $|x-y|>r$). We consider the `wizard hat'~\cite{coombes05b,kilpatrick10c}, for explicit calculations:
\begin{align}
w_u(x) = (1- |x|)\e^{-|x|}, \label{wu}
\end{align}
so $w(x)>0$ when $|x|<1$ and $w(x)<0$ when $|x|>1$. The Heaviside nonlinearity in Eq.~(\ref{pfield}), 
\begin{align*}
H(u-\theta_u) = \left\{ \begin{array}{cl} 1 & \ : u>\theta_u, \\ 0 & \ : u<\theta_u, \end{array} \right.
\end{align*}
models thresholding that converts synaptic input to output firing rate~\cite{bressloff12}. More nuanced firing rate functions are possible but complicate calculations and do not substantially alter the qualitative results~\cite{coombes10}. The impact of velocity inputs is modeled by the final term in Eq.~(\ref{pfield}), which propagates stationary bump solutions in the direction of the time-dependent velocity vector $v (t)$~\cite{zhang96}. This feature is ensured by the form of the weight function $w'(x)$ in the convolution, which results in a translation of the bumps, as we will demonstrate in Section \ref{lowdim}. Given the weight function Eq.~(\ref{wu}) we have chosen, then
\begin{align*}
w_u'(x) = - \sign (x) \left( 2 - |x| \right) \e^{- |x|},
\end{align*}
where a jump discontinuity arises at $x=0$, which is mollified by the integration in Eq.~(\ref{pfield}). \\
\vspace{-3mm}

\noindent
{\bf Memory layer.} As we have discussed, IOR biases organisms' search strategy, so they are less likely to visit locations that have already been scanned~\cite{klein88,horowitz03}. We account for neural hardware capable of storing visited locations over appreciable periods of time (e.g., at least several seconds~\cite{wang10}).
Since IOR elicits cortical activity in oculomotor programming areas~\cite{sapir99,lepsien02,mayer04}, we describe memory of previously visited locations with a neural activity variable $q(\x,t)$ on $\x \in (- \infty, \infty)$ and $t \in [0, \infty)$ spanning all searchable locations:
\begin{align}
q_t &= -q + w_q\;\bar{*}\;H(q-\theta_q)+ w_p*H(u-\theta_u). \label{mfield}
\end{align}
Recurrent coupling in Eq.~(\ref{mfield}) is described by the local synaptic weight kernel $w_q(x,y)$, with spatial heterogeneity, so that $w_q(x,y) \not\equiv w_q(x - y)$, and $w_q\;\bar{*}\;H(q-\theta_q) = \int_{- \infty}^{\infty} w_q(x,y) H(q(y,t)- \theta_q) \d y$. The scale of the heterogeneity may be set by environmental landmarks~\cite{stemmler95} or the spatial extent of receptive fields or hypercolumns in the visual cortex~\cite{angelucci06}. Heterogeneity can stabilize activity, so that excitation does not cause it to propagate indefinitely~\cite{bressloff01,coombes11,avitabile15}. We will typically assume such spatial heterogeneity is periodic, specifically given by a cosine-modulated exponential:
\begin{align}
w_q(x,y) : =& \left[ 1 + \sigma w_h(y) \right] \bar{w}_q(x-y) \nonumber\\
=& \left[ 1 + \sigma \cos (n y) \right] \cdot \frac{\e^{-|x-y|}}{2}, \label{wq}
\end{align}
so the weight function is a homogeneous kernel $\bar{w}_q(x-y)$ modulated by a periodic heterogeneity $w_h(y)$. Varying the amplitude $\sigma$ of the heterogeneity increases the parameter range for which traveling wave solutions are pinned, yielding stationary activity. For perturbative results, we assume $0 < \sigma \ll 1$. The parameter $n$ sets the number of stable conformations of wave fronts over the length $2 \pi$. Input from the position layer is weak, as compared with local recurrent connections, and excitatory,
\begin{align}
w_p(x) = I_0 \frac{\alpha \e^{-\alpha|x|}}{2},   \label{wp}
\end{align}
so typically we use the assumption $0 < I_0 \ll 1$. Varying the inverse spatial scale $\alpha$ shapes the precision with which position information is sent from the position layer to the memory layer. We found changing $\alpha$ had minimal impact on our qualitative results, so we mostly set $\alpha = 1$.

Velocity $v(t)$ is a nonlinear combination of a baseline program of velocity $v_o(t)$, which would tend to guide the agent with no memory-based feedback, and a control function $\chi (u,q)$, which depends on both the position $u(x,t)$ and the memory $q(x,t)$ layer activities. We discuss different instantiations of this control in in Section \ref{perform}, where we examine performance the model in two different types of search tasks.  \\
\vspace{-3mm}

\section{Stationary solutions}
\label{1D}

We begin by studying the existence and stability of stationary solutions the neural field model, Eq.~(\ref{pfield}) and (\ref{mfield}), in the case of no velocity input ($v(t) \equiv 0$) and strong input position input ($I_0$ arbitrary) and heterogeneity ($\sigma$ arbitrary). Understanding these stationary solutions, and specifically the effect of the position layer on the memory layer, will provide us with insight needed to project the neural field to a low-dimensional model that can be more easily analyzed. Subsequently, we will consider the effect of closed-loop control on the velocity input based upon the combination of position and memory layer activities. \\
\vspace{-3mm}

\begin{figure*}
\begin{center} \includegraphics[width=16cm]{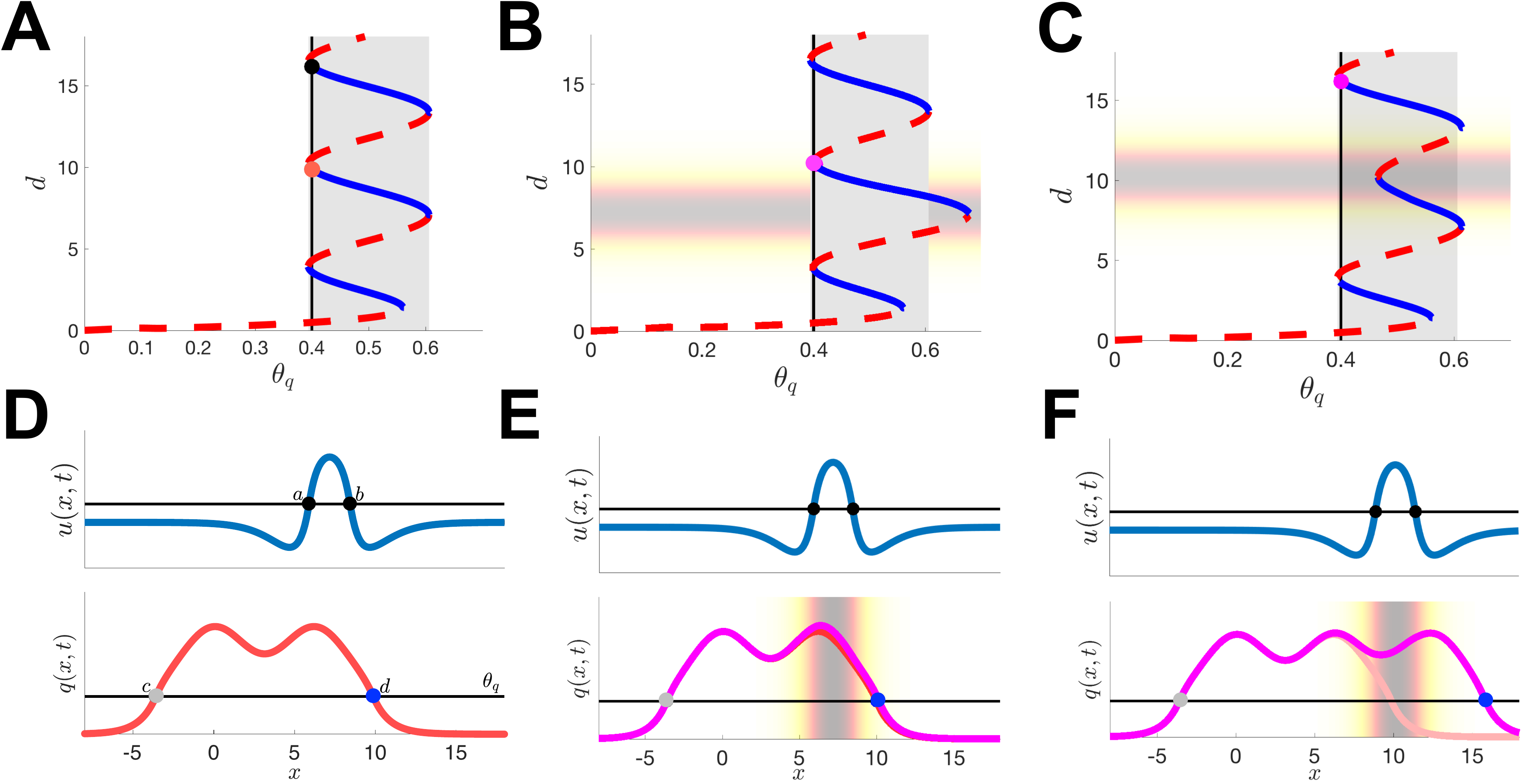} \end{center}
\vspace{-4mm}
\caption{Dependence of stationary solutions on the input from the position layer. {\bf A.} For no position layer input ($ I_0 = 0$), there is a range of memory layer threshold values $\theta_q$ (grey region), for which multiple stable (blue solid) and unstable (red dashed) standing front solutions exist. The branches ``snake" back and forth, turning at saddle-node (SN) bifurcations (See also \cite{avitabile15}). The left interface $c$ is bounded to be between $-2 \pi$ and $0$ (see main text), and we solve Eq.~(\ref{iface}) to obtain the both the left $c$ and right interface (see {\bf D} for reference). {\bf B.} Weak input ($ I_0 = 0.1$, background gradient) from a bump centered at $\Delta_u = 7.184$ (near 3rd SN) shifts branches near the 3rd SN, but weakly affects other branches. A stable solution near $d \approx 10$  for $\theta_q = 0.4$ still remains (magenta dot). {\bf C.} Weak input from bump centered at $\Delta_u = 10.184$ (near 4th SN) shifts branches near $d \approx 10$, so nearby solutions for $\theta_q = 0.4$ vanish. {\bf D,E,F.} Stationary solution profiles associated with red dot in {\bf A} (panel {\bf D}); magenta dot in {\bf B} (panel {\bf E}); and magenta dot in {\bf C} (panel {\bf F}). Solution from {\bf D} is shown for reference in panels {\bf E} and {\bf F}. Other parameters are $\theta_u = 0.2$, $\theta_q = 0.4$, $\alpha = 1$, $n=1$, $ \sigma = 0.3$.}
\label{fig3_bifdiag}
\end{figure*}

\noindent
{\bf Existence.} We begin with the stationary equations associated with Eq.~(\ref{pfield}) and (\ref{mfield}) following these assumptions, finding
\begin{subequations}  \label{stateqns}
\begin{align}
U(x) &= w_u*H(U(x)- \theta_u), \label{stateqn1} \\
Q(x) &= w_q\;\bar{*}\;H(Q(x) - \theta_q) + w_p*H(U(x) -\theta_u).
\end{align} 
\end{subequations}
Since the nonlinearities are step functions, we can simplify the stationary equations, Eq.~(\ref{stateqns}), by constraining the form of their solutions. In particular, we look for solutions with simply-connected active regions in both layers: a bump in the position layer and a stationary front in the memory layer. These assumptions lead to the following conditions: $U(x) > \theta_u$ for $x \in (a, b)$ and $Q(x) > \theta_q$ for $x \in (c,d)$. Often in studies of wave solutions to neural fields, translation invariance is used to project out one of the wave interface parameters~\cite{bressloff12}. However, it is important to note here that while the position layer $U(x)$ does have translation symmetry, the memory layer $Q(x)$ does not. Thus, to fully characterize qualitatively different solutions, we must explore all possible vector solutions $(a,b,c,d)$ to the resulting reduced equations
\begin{subequations}  \label{stateqns2}
\begin{align}
U(x) &= \int_{a}^{b} w_u(x-y) \d y, \\
Q(x) &= \int_{c}^{d} w_q(x,y) \d y + \int_{a}^{b} w_p(x-y) \d y.
\end{align} 
\end{subequations}
For specific choices of the weight kernels, the integrals in Eq.~(\ref{stateqns2}) can be computed explicitly, easing the identification of solutions. In particular, we use the kernels defined in Eqs.~(\ref{wu}), (\ref{wq}), and (\ref{wp}) and evaluate integrals to find that stationary solutions take the form
\begin{subequations} \label{statsoln}
\begin{align}
U(x) &= (x-a) \e^{-|x-a|} - (x-b) \e^{-|x-b|}, \\
Q(x) &= {\mc F}(x;c,d) + {\mc P}(x;a,b),
\end{align}
\end{subequations}
where ${\mc F}$ and ${\mc P}$ are defined piecewise, according to the active regions $R_q \equiv (c,d)$ and $R_u \equiv (a,b)$ as 
\begin{align*}
{\mc F}(x;c,d) = \left\{ \begin{array}{cc} {\mc M}_+(x,d) - {\mc M}_+(x,c), & x\in[d,\infty], \\
\D {\mc C}(x)- {\mc M}_+(x,c) - {\mc M}_-(x,d), & x \in [c,d], \\
\D {\mc M}_-(x,c) - {\mc M}_-(x,d), & x\in[-\infty,c], \end{array} \right.
\end{align*}
where ${\mc C}(x) = 1 + \sigma \frac{\cos (nx)}{n^2+1}$,
\begin{align*}
{\mc M}_{\pm}(x,y) = \frac{\e^{\mp (x-y)}}{2} \left[ 1 + \sigma \frac{\cos (ny) \pm n \sin (ny)}{n^2 +1} \right],
\end{align*}
and
\begin{align*}
{\mc P}(x;a,b) =& \frac{I_0}{2} \left[ {\rm sign} (b-x) \left( 1 - \e^{- \alpha |x-b|} \right) \right. \\
& \hspace{11mm} \left. + {\rm sign} (x-a) \left( 1 - \e^{- \alpha |x-a|} \right)  \right],
\end{align*}
where $\sign (z) = \pm 1$ if $z \gtrless 0$ and $\sign (0) = 0$. Together with the threshold conditions $U(a) = U(b) = \theta_u$ and $Q(c) = Q(d) = \theta_q$, we have implicit equations for the interface locations:
\begin{subequations}  \label{iface}
\begin{align}
\theta_u &= (b-a) \e^{a-b},  \\
\theta_q & =  {\mc M}_-(c,c) - {\mc M}_-(c,d) +  {\mc P}(c;a,b),  \\
\theta_ q & = {\mc M}_+(d,d) - {\mc M}_+(d,c) +  {\mc P}(d;a,b).
\end{align}
\end{subequations}
Note, a degeneracy arises in the equation for the bump interfaces $a$ and $b$, due to the translation symmetry of the position equation, Eq.~(\ref{pfield}). Since Eq.~(\ref{iface}) contains a mixture of transcendental functions, we do not expect to be able to solve explicitly for vector solutions $(a,b,c,d)$. Thus, we will employ a nonlinear root-finder in order to construct associated bifurcation diagrams.

We now demonstrate the mechanism by which fronts are propagated in Eq.~(\ref{mfield}), via input from the position layer, Eq.~(\ref{pfield}). This analysis uses bifurcation diagrams associated with stationary solutions, but in Section \ref{lowdim} we approximate the dynamics of the bump and front interfaces to obtain a low-dimensional system for the motion of the patterns in each layer. Bifurcation curves and stationary solutions of the model Eq.~(\ref{pfield}) and (\ref{mfield}) are shown in Fig. \ref{fig3_bifdiag}. Nonlinear root-finding applied to Eq.~(\ref{iface}) is used to compute the bifurcation curves, but the stability will be determined by a linear analysis below. To clearly display solution curves, we have bounded the left interface $c$ of the front solution between $[- 2\pi,0]$. Note, similar bifurcation curves would be obtained by bounding $c \in [- 2(m+1) \pi, -2 m \pi]$ for any positive integer $m$. The location of the left interface $c$ only marginally affects the right interface, since interactions between the interfaces are described by the function $\e^{c-d}$ in Eq.~(\ref{iface}), which will typically be small. In \cite{avitabile15}, this was addressed by plotting bifurcation diagrams showing the dependence of the width $L_f = d-c$ of the front, rather than the right interface $d$. For our purposes, it is more instructive to track how $d$ changes with the location of the bump in the position layer.

Solutions' dependence on the memory layer threshold $\theta_q$ and input from the position layer is shown in Fig. \ref{fig3_bifdiag}A--C. The case of no input ($ I_0 \equiv 0$ in Eq.~(\ref{iface})) is shown in Fig. \ref{fig3_bifdiag}A. Note the metastability of solution in the grey shaded region. Advancing the front interface to subsequent stable branches is the main mechanism by which previously visited locations are stored by the network. When input from the position layer is applied to the memory layer (Fig. \ref{fig3_bifdiag}B,C), it warps the solution curves in the vicinity of the excitation. This can result in the annihilation of stable solutions at lower values of $\theta_q$ (Fig. \ref{fig3_bifdiag}C). We plot profiles in Fig. \ref{fig3_bifdiag}D--F, demonstrating how solutions are identified with their threshold intersection points. Note input from the position layer is not sufficient to destabilize the input-free solution in Fig. \ref{fig3_bifdiag}E, but is in Fig. \ref{fig3_bifdiag}F since the input is slightly ahead of the front interface. This is the mechanism by which the memory layer's front is propagated, once moving bumps in the position layer are considered: The bump must be ahead of the front interface to propagate it forward.

\begin{figure*}
\begin{center} \includegraphics[width=17cm]{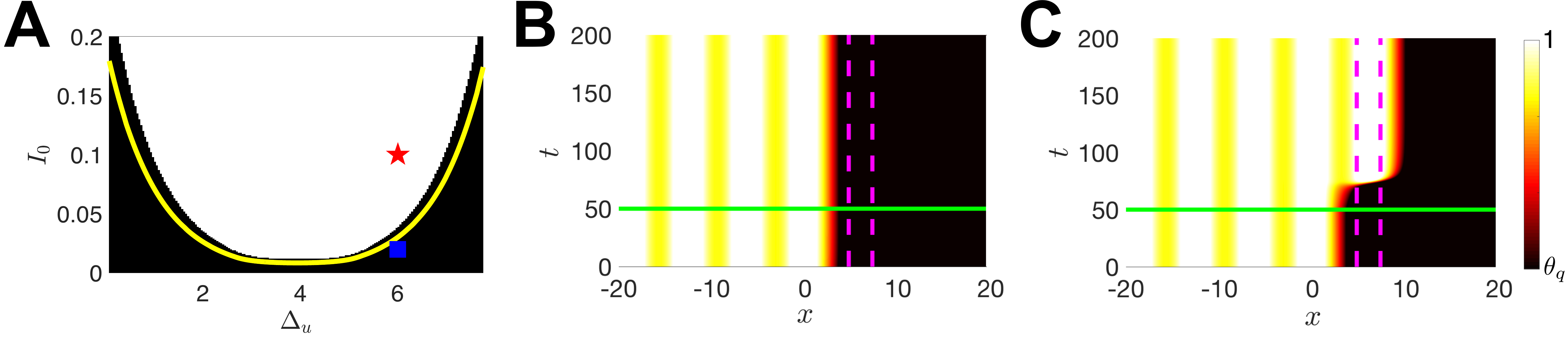}  \end{center}
\vspace{-4mm}
\caption{Phase diagram showing the impact of the bump location $\Delta_u$ and input amplitude $I_0$ on the movement of a nearby front interface. {\bf A.} Partition of $(I_0, \Delta_u)$ parameter space into regions where front propagates forward (white) and where it does not (black). Line shows analytical approximation from Eq.~(\ref{critI}). Mismatch arises due to subtle discretization errors in the numerical scheme for solving the full system. {\bf B,C.} Numerical simulations of the full network, Eq.~(\ref{pfield}) and (\ref{mfield}) with a stationary bump centered at $\Delta_u = 6$ (interfaces given by dashed lines). The position input amplitude is switched from $I_0=0$ to $I_0 = 0.02 $ in panel {\bf B} (blue square in {\bf A}) and $I_0=0.1$ in panel {\bf C} (red star in {\bf A}) at $t=50$ (light line). The front only propagates for large enough $I_0$ (panel {\bf C}). Other parameters are $\theta_u = 0.2$, $\theta_q = 0.4$, $\sigma = 0.3$, $n=1$, and $\alpha = 1$.}
\label{fig4_phase}
\end{figure*}

Next, we take a closer look at the bifurcation that occurs by increasing the strength $I_0$ of the input from the position layer to the memory layer. In particular, we consider a one-sided front, as this a fairly accurate approximation to the case $d-c \gg 1$, and we will utilize this observation in our low-dimensional system we derive in Section \ref{lowdim}.\\
\vspace{-3mm}

\noindent
{\bf One-sided front.} Terms involving $\e^{c-d}$ will tend to be quite small even for a modest difference between the two front interfaces (e.g., $\e^{-10} \approx 4.54 \times 10^{-5}$). Thus, we consider the case where $c-d$ is sufficiently large as to ignore the exponentially small term $\e^{c-d}$, and focus specifically on using Eq.~(\ref{iface}) to solve for the right interface $d$. In this case, we can write
\begin{subequations}  \label{iface2}
\begin{align}
\theta_u &= (b-a) \e^{a-b},  \label{iface2a} \\
\theta_ q & = \frac{1}{2}  + \frac{ \sigma \cos (nd) + n  \sigma \sin (nd) }{2(n^2+1)} +  {\mc P}(d;a,b),  \label{iface2b} 
\end{align}
\end{subequations}
so that Eq.~(\ref{iface2a}) and (\ref{iface2b}) can be solved in sequence to obtain bifurcation curves for $d$, which will be very similar to Fig. \ref{fig3_bifdiag}, except for small differences arising for low values of $d$.
As we also demonstrated in Fig. \ref{fig3_bifdiag}C, increasing the strength of the input $I_0$ from the position layer to the memory layer can lead to the annihilation of the pair of stable/unstable solutions via a saddle-node (SN) bifurcation. We will now discuss how to identify this curve of SN bifurcations in the reduced Eq.~(\ref{iface2}).

Fixing the threshold $\theta_q$, we can identify the critical $I_0$ for which a SN bifurcation occurs by simultaneously looking for the interface location $d^c$ and $I_0^c$ at which an extremum of the right-hand-side of Eq.~(\ref{iface2b}) occurs. In essence, this identifies the point at which the bend of the bifurcation curves in Fig. \ref{fig3_bifdiag} cross through a threshold value $\theta_q$ due to an increase of the input $I_0$. This requires, first, that Eq.~(\ref{iface2b}) is satisfied for $d = d^c$. Additionally, we require that the derivative of the right-hand side Eq.~(\ref{iface2b}) with respect to $d$ is zero since the SN bifurcation occurs at a critical point of the solution curve. Regardless of the location of the bump in the position layer (parameterized by $a$ and $b$), this condition is given by the equation:
\begin{align*}
\frac{n  \sigma}{n^2+1} \left[ n \cos (nd) - \sin (nd) \right] \hspace{26mm} &\\
+ \frac{ \alpha I_0}{2} \left[ \e^{ - \alpha |d-a|} - \e^{-\alpha |d-b|} \right] &= 0,
\end{align*}
which can be solved explicitly for the critical input strength $I_0^c$ in terms of the interface location $d^c$ at the bifurcation:
\begin{align}
I_0^c = \frac{2 n \sigma}{\alpha (n^2+1)} \cdot \frac{\sin (nd^c) - n \cos (nd^c)}{\D \e^{ - \alpha |d^c - a|} - \e^{- \alpha |d^c-b|}}.  \label{critI}
\end{align}
Plugging Eq.~(\ref{critI}) into Eq.~(\ref{iface2b}), we obtain the following implicit equation for the critical location of the interface, given the critical input $I_0^c$:
\begin{align}
\theta_q & =  \frac{1}{2} + \sigma \frac{\cos (nd^c) + n \sin (n d^c)}{2(n^2+1)} +  \frac{n \sigma}{\alpha (n^2+1)}  \label{critdc} \\
& \hspace{20mm} \times \frac{\sin (nd^c) - n \cos (nd^c)}{\D \e^{ - \alpha |d^c - a|} - \e^{- \alpha |d^c-b|}} \frac{{\mc P}(d^c;a,b)}{I_0}. \nonumber
\end{align}
Eq.~(\ref{critdc}) further simplifies in the case where the input is ahead of the interface $d^c< a<b$, so that
\begin{align*}
\frac{2 \alpha}{\sigma} (n^2+1) \left( \theta_q - \frac{1}{2} \right) &= \left( \alpha  - 2 n^2 \right) \cos (nd^c) \\
& \hspace{15mm} + (\alpha +2) n \sin (nd^c),
\end{align*}
which can be solved explicitly
\begin{align*}
d^c &= \frac{2}{n} \left[ \tan^{-1} \left( \frac{\sqrt{{\mc A}^2 + {\mc B}^2 - {\mc C}^2} + {\mc B}}{{\mc A}+{\mc C}} \right) + m \pi \right],
\end{align*}
for $m \in \mathbb{Z}$, where ${\mc A} =  \alpha  - 2 n^2$, ${\mc B} = (\alpha + 2)n$, and ${\mc C} = \frac{2 \alpha}{\sigma} (n^2+1) \left( \theta_q - \frac{1}{2} \right)$. A similar set of explicit solutions can be obtained for the case $a<b < d^c$, so that
\begin{align*}
d^c &= \frac{2}{n} \left[ \tan^{-1} \left( \frac{\sqrt{{\mc A}^2 + {\mc B}^2 - {\mc C}^2} - {\mc B}}{{\mc C} - {\mc A}} \right) + m \pi \right].
\end{align*}
We cannot solve the case $a< d^c < b$ explicitly, but it can easily be evaluated using numerical root finding.

We compare our analytical results to numerical simulations in Fig. \ref{fig4_phase}. In particular, we study the input strength $I_0$ necessary to advance the location of the front in the memory layer to the next stable branch of solutions (See Fig. \ref{fig3_bifdiag}B,C for further illustration). As the location $\Delta_u$ of the bump in the position layer is varied, the critical input strength $I_0^c$ at which the front in the memory layer advances varies nonmonotonically (Fig. \ref{fig4_phase}A). At an intermediate value of the bump location, $I_0^c$ obtains a minimum. Thus, activity in the memory layer propagates best when the bump is slightly advanced as compared to the front location. Our results match well with numerical simulations, which predict a similar relationship between the location of the bump $\Delta_u$ and the critical input strength $I_0^c$. We demonstrate examples of the front dynamics in the full system's, Eqs.~(\ref{pfield}) and (\ref{mfield}), in Fig. \ref{fig4_phase}B,C. When $I_0<I_0^c$, the front does not advance to the next stable branch, but the front does advance if $I_0 > I_0^c$.


Next, we examine the linear stability of the solutions computed above. In particular, we expect to find zero eigenvalues associated with the front solutions $Q(x)$ at the SN bifurcation points. Such solutions indicate boundaries at which $I_0$ can be changed to propagate the front from one stable branch to the next. \\
\vspace{-3mm}

\begin{figure*}
\begin{center} \includegraphics[width=13cm]{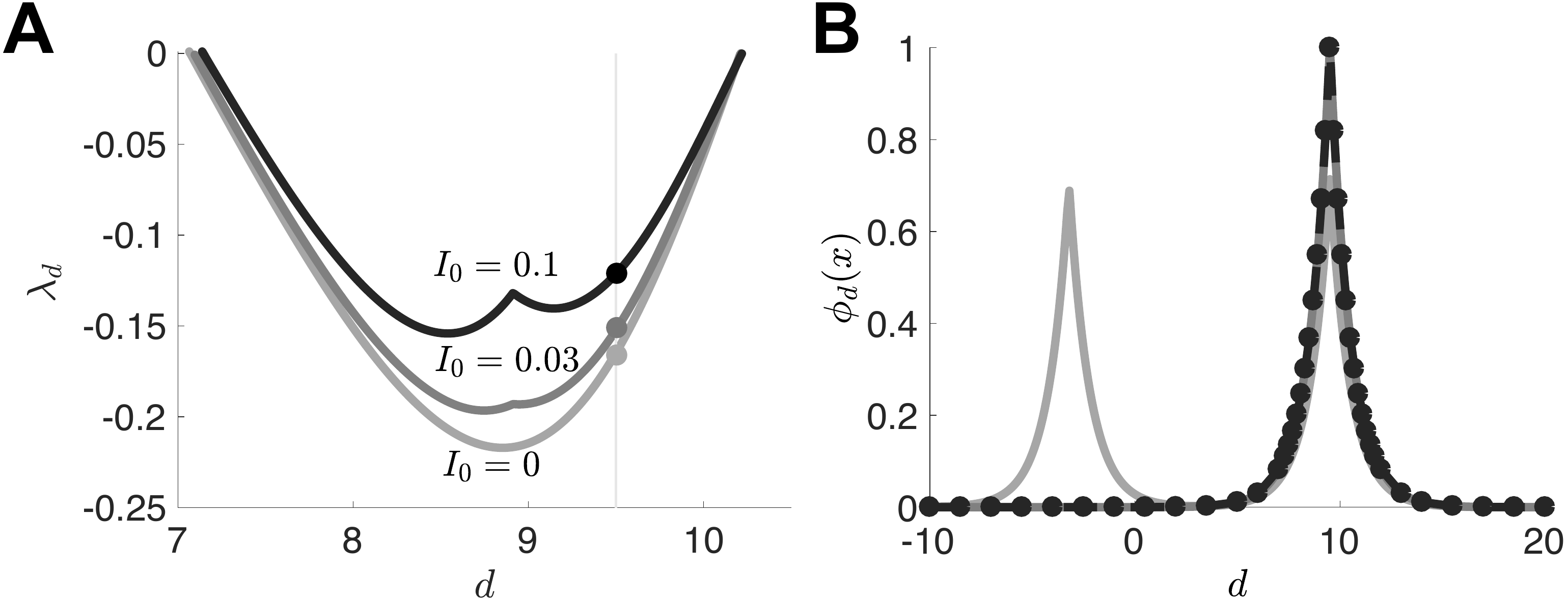}  \end{center}
\vspace{-4mm}
\caption{Eigensolution for perturbations near the right interface of the front at $x=d$. {\bf A.} Eigenvalue varies nonmonotonically for different values of $\theta_q$ (correspondingly plotted versus $d$) along the stable branch corresponding to those $d$ values. As input amplitude $I_0$ is increased, the range of $d$ values decreases and $\lambda_d : = \lambda_+$, Eq.~(\ref{qeig}), moves closer to zero. Cusp arises at boundary of input (bump in position layer is centered at $\Delta_u= 10.184$). {\bf B.} Eigenfunction $\phi_d(x)$ associated with perturbations of the right interface, $x=d$, defined in Eq.~(\ref{linsys3}). Peak near $x=d$ increases as the input amplitude is increased $I_0$. Shades correspond to input strength $I_0$ as in {\bf A.} Other parameters are $\sigma = 0.3$, $\alpha = 1$, $n = 1$, $c \in [-2 \pi ,0]$, and $\theta_u = 0.2$.}
\label{fig5_eigs}
\end{figure*}

\noindent
{\bf Linear stability.} The stability of stationary solutions, defined by Eq.~(\ref{iface}), can be determined by examining the evolution of spatiotemporal perturbations, $\ep (\psi (x,t), \phi (x,t))$, where $0< \ep \ll 1$. We examine linearized equations associated with the perturbed solutions $u(x,t) = U(x) + \ep \psi (x,t)$ and $q(x,t) = Q(x) + \ep \phi (x,t)$. Plugging these into the original evolution Eq.~(\ref{pfield}) and (\ref{mfield}), for $v(t) \equiv 0$, and truncating to ${\mc O}(\ep)$, we obtain
\begin{align}
\psi_t &= - \psi  + w_u*\left[ H'(U - \theta_u) \psi \right],  \label{linsys1} \\
\phi_t &= - \phi  + w_q*\left[ H'(Q - \theta_q) \phi \right] + w_p*\left[ H'(U - \theta_u) \phi  \right]. \nonumber
\end{align} 
The spectrum of the associated linear operator is found by examining the evolution of the separable solutions $\psi (x,t) = \e^{\lambda t} \psi (x)$ and $\phi (x,t) = \e^{\lambda t} \phi (x)$. Furthermore, the convolutions in Eq.~(\ref{linsys1}) are localized, since they involve derivatives of $H(U(x) - \theta_u) = H(x-a) - H(x-b)$ and $H(Q(x) - \theta_q) = H(x-c) - H(x-d)$, which are
\begin{align*}
\delta (x-a) - \delta (x-b) &=  \frac{\d H(U - \theta_u)}{\d x} = H'(U - \theta_u) U', \\
\delta (x-c) - \delta (x-d) &= \frac{\d H(Q - \theta_q)}{\d x} = H'(Q - \theta_q) Q',
\end{align*}
which can be rearranged to find
\begin{subequations} \label{Hdiff}
\begin{align}
H'(U(x) - \theta_u) &= \frac{1}{|U'(a)|} \left[ \delta (x-a) + \delta (x-b) \right], \\
H'(Q(x) - \theta_q) &= \frac{\delta (x-c)}{|Q'(c)|} + \frac{\delta (x-d)}{|Q'(d)|},
\end{align}
\end{subequations}
where
\begin{align*}
U' &= w_u(x-a) - w_u(x-b), \\
Q' &= \int_c^d \frac{\d w_q(x,y)}{\d x}  \d y + w_p(x-a) - w_p(x-b).
\end{align*}
We can assume even symmetry of the bump solution $U(x)$, but not the front solution $Q(x)$. Applying the identities in Eq.~(\ref{Hdiff}) to Eq.~(\ref{linsys1}) along with separability, we obtain the following system for the spectrum of the underlying linear operator:
\begin{subequations} \label{linsys3}
\begin{align}
(\lambda + 1) \psi &= \gamma_a \left[ w_u(x-a) \psi (a) + w_u(x-b) \psi (b) \right], \label{linsys3a} \\
(\lambda +1) \phi &= \gamma_a \left[ w_p(x-a) \psi (a) + w_p(x-b) \psi (b) \right] \label{linsys3b} \\
& \hspace{9mm} + \gamma_c w_q(x,c) \phi (c) + \gamma_d w_q(x,d) \phi (d), \nonumber
\end{align}
\end{subequations}
where $\gamma_a^{-1} = |U'(a)|$, $\gamma_c^{-1} = |Q'(c)|$, and $\gamma_d^{-1} = |Q'(d)|$. There are two classes of solution to Eq.~(\ref{linsys3}). First, all solutions $\psi (a) = \psi (b) = \phi (c) = \phi (d) = 0$ lie in the essential spectrum and $\lambda = -1$, which contributes to no instabilities. Solutions that do not satisfy the condition $\psi (a) = \psi (b) = \phi (c) = \phi (d) = 0$ can be classified by the vector $(\psi (a), \psi (b), \phi (c), \phi (d))$. In this case, the functions $(\psi (x), \phi (x))$ are fully specified by their values at $(a,b,c,d)$. This leads to the linear system
\begin{align}
( \lambda +1) \psi (a) &= \gamma_a \left[ w_u (0) \psi (a) + w_u(b-a) \psi (b) \right], \label{fourdim} \\
( \lambda + 1) \psi (b) & = \gamma_a \left[ w_u(b-a) \psi (a) + w_u(0) \psi (b) \right], \nonumber \\
(\lambda + 1) \phi (c) &= \gamma_a \left[ w_p(c-a) \psi (a) + w_p(c-b) \psi (b) \right] \nonumber \\
& \hspace{8mm} + \gamma_c w_q(c,c) \phi (c) + \gamma_d w_q(c,d) \phi (d), \nonumber \\
(\lambda +1) \phi (d) &= \gamma_a \left[ w_p(d-a) \psi (a) + w_p(d-b) \psi (b) \right] \nonumber \\
& \hspace{8mm} + \gamma_c w_q(d,c) \phi (c) + \gamma_d w_q(d,d) \phi (d). \nonumber
\end{align}
Since Eq.~(\ref{fourdim}) is in block triangular form, the eigenvalue problem can be separated in to diagonal blocks~\cite{silvester00}. The upper left block yields a two-by-two eigenvalue problem with eigenvectors $(\psi (a), \psi(b))$~\cite{amari77,ermentrout98,bressloff12}:
\begin{align*}
( \lambda +1) \psi (a) &= \gamma_a \left[ w_u (0) \psi (a) + w_u(b-a) \psi (b) \right], \\
( \lambda + 1) \psi (b) & = \gamma_a \left[ w_u(b-a) \psi (a) + w_u(0) \psi (b) \right],
\end{align*}
leading to the following characteristic equations for the associated eigenvalues:
\begin{align*}
{\mc U}( \lambda ) &= \left| \begin{array}{cc} \lambda + 1 - \gamma_a w_u(0) & - \gamma_a w_u(b-a) \\
- \gamma_a w_u(b-a) & \lambda + 1 - \gamma_a w_u(0) \end{array} \right| \\
&= \lambda \left( \lambda - \frac{2 w_u(b-a)}{w_u(0) - w_u(b-a)} \right) = 0.
\end{align*}
The other two-by-two system corresponds to eigenvectors of the form $(\phi (c), \phi (d))$:
\begin{subequations}. \label{fronteprob}
\begin{align}
(\lambda + 1) \phi (c) &=  \gamma_c w_q(c,c) \phi (c) + \gamma_d w_q(c,d) \phi (d), \\
(\lambda +1) \phi (d) &=  \gamma_c w_q(d,c) \phi (c) + \gamma_d w_q(d,d) \phi (d),
\end{align}
\end{subequations}
with corresponding characteristic equation
\begin{align*}
{\mc Q}(\lambda) &= \left| \begin{array}{cc} \lambda + 1 - \gamma_c w_q(c,c) & - \gamma_d w_q(c,d) \\
- \gamma_c w_q(d,c) & \lambda + 1 - \gamma_d w_q(d,d) \end{array} \right| \\
& = \lambda^2 + {\mc Q}_1 \lambda + {\mc Q}_0 = 0, 
\end{align*}
where ${\mc Q}_0 = (\gamma_c w_q(c,c) - 1)(\gamma_d w_q(d,d) -1) - \gamma_c \gamma_d w_q(c,d) w_q(d,c)$ and ${\mc Q}_1 = 2 - \gamma_c w_q(c,c) - \gamma_d w_q(d,d)$. Clearly, the roots of ${\mc U}( \lambda ) = 0$ are $\lambda_0 = 0$ and $\D \lambda_w =  2 w_u(b-a)/ (w_u(0) - w_u(b-a))$, the typical stability classification of bumps in neural fields with Heaviside firing rates~\cite{amari77,bressloff12}. The zero eigenvalue indicates the translation symmetry of the bump, and the generically nonzero eigenvalue $\lambda_w$ represents the stability of the bump in response to width perturbations, determined by the sign of $w_u(b-a)$. We are interested in the linear stability characterized by ${\mc Q}(\lambda) = 0$. These eigenvalues can be determined explicitly assuming $(a,b,c,d)$ and $Q'(x)$ are known by applying the quadratic formula
\begin{align}
\lambda_{\pm} &= \frac{1}{2} \left[ -{\mc Q}_1 - \sqrt{{\mc Q}_1^2 - 4 {\mc Q}_0} \right],  \label{qeig} 
\end{align}
Neutral stability of the front occurs when ${\rm Re} \lambda_+ = 0$. Past work showed these are SN bifurcations~\cite{avitabile15}, so we expect $\lambda_+ = 0$. Placing this condition on Eq.~(\ref{qeig}) yields ${\mc Q}_0 = 0$, and
\begin{align*}
(\gamma_c w_q(c,c) -1) (\gamma_d w_q(d,d) -1)  = \gamma_c \gamma_d w_q(c,d) w_q(d,c).
\end{align*}
For front interfaces that are far apart $d-c \gg 1$, $|w_q (c,d)|, |w_q(d,c)| \ll 1$. These terms scale exponentially with the distance $d-c$, so their product will be smaller, and we approximate Eq.~(\ref{fronteprob}) as a diagonal system with eigenvalues $\lambda_d : = \lambda_+  = \gamma_d w_q(d,d)  -1$ and $\lambda_c : = \lambda_- = \gamma_c w_q(c,c) - 1$. Focusing on bifurcations that emerge at the right interface near $x = d$, we expect the SN bifurcation to occur when $\lambda_d = 0$ or $w_q(d,d) = |Q'(d)|$. This is identical to the condition we derived above for the location of SN bifurcations for a single-interface front.

We solve for the eigenvalue $\lambda_d$ using the formula Eq.~(\ref{qeig}) and plot in Fig. \ref{fig5_eigs}A, showing the dependence of this eigenvalue for a specific front interface value $d$. As expected, the eigenvalue becomes zero at the endpoints of the stable branch, annihilating in a SN bifurcation. The associated eigenfunction is determined by plugging $\lambda_d = \lambda_+$ into Eq.~(\ref{fourdim}), solving the linear system for the degenerate eigenvector $(\psi (a), \psi (b), \phi (c), \phi (d))$, and using the full linear system Eq.~(\ref{linsys3}) to determine the shape of $(\psi (x), \phi (x))$. The result is plotted in Fig. \ref{fig5_eigs}B, showing perturbations of this form shift the location of the right front interface to the right.

With knowledge of the mechanism by which the position layer $u(x,t)$ moves the memory layer front $q(x,t)$, we now derive low-dimensional approximations of the bump and front motion. Our interface calculations track the location of the bump $\Delta_u(t)$ as well as the left and right locations of the memory front $\Delta_+(t)$ and $\Delta_-(t)$ using a system of three nonlinear differential equations.

\section{Interface equations}
\label{lowdim}

We now derive interface equations for the position layer $u(x,t)$ and the front layer $q(x,t)$, starting with mild assumptions on the parameters and dynamics of the activity in each layer. Strong assumptions of weak heterogeneity and inputs will be used to simplify the form of our derived interface equations. Interface equations reduce the dimensionality of our system due to the Heaviside form of the nonlinearities in Eq.~(\ref{pfield}) and (\ref{mfield}), so that the threshold crossing points $u(x,t) = \theta_u$ and $q(x,t) = \theta_q$ largely determine the dynamics of the full system. Several authors extended the seminal work of Amari (1977)~\cite{amari77}, who developed interface methods for analyzing bump stability in neural fields, to handle more complicated dynamics like fronts in heterogeneous networks~\cite{coombes11} and two-dimensional domains~\cite{coombes12}.

To start, we define the active regions of both layers, $A_u(t) = \{ x | u(x,t) > \theta_u \}$ and $A_q(t) = \{ x | q(x,t) > \theta_q \} $ where the output of the firing rate nonlinearities will be nonzero, allowing us to rewrite Eq.~(\ref{pfield}) and (\ref{mfield}) as 
\begin{align}
u_t +u &= \int_{A_u(t)} w_u(x-y) \d y - v(t) \int_{A_u(t)} w_u'(x-y) \d y, \nonumber \\
q_t +q &= \int_{A_q(t)} w_q(x,y) \d y + \int_{A_u(t)} w_p(x-y) \d y, \label{indyn1}
\end{align}
and since we expect the active regions to be simply-connected, we specify
\begin{align*}
A_u(t) = (x_-(t), x_+(t)), \hspace{4mm} A_q(t) = (\Delta_-(t),\Delta_+(t)).
\end{align*}
Assuming continuity of $u(x,t)$ and $q(x,t)$, the boundaries of $A_{u,q}(t)$ describe the interfaces of the bump and front. Thus, we write the dynamic threshold equations
\begin{align}
u(x_{\pm}(t) ,t) = \theta_u, \hspace{4mm} q(\Delta_{\pm}(t),t) = \theta_q. \label{thresh1}
\end{align}
Differentiating Eq.~(\ref{thresh1}) with respect to $t$, we find
\begin{subequations}   \label{indyn2}
\begin{align}
\alpha_{\pm}(t) \frac{\d x_{\pm}}{\d t} + \frac{\pd u(x_{\pm}(t),t)}{\pd t} &= 0, \\
\beta_{\pm} (t) \frac{\d \Delta_{\pm}}{\d t} + \frac{\pd q(\Delta_{\pm}(t),t)}{\pd t} &= 0,
\end{align}
\end{subequations}
where we have defined
\begin{align*}
\alpha_{\pm} (t) = \frac{\pd u(x_{\pm}(t),t)}{\pd x}, \hspace{4mm} \beta_{\pm}(t) = \frac{\pd q(\Delta_{\pm}(t),t)}{\pd x}.
\end{align*}
We obtain differential equations for the dynamics of the interfaces $x_{\pm}(t)$ and $\Delta_{\pm} (t)$ by rearranging Eq.~(\ref{indyn2}) and substituting Eq.~(\ref{indyn1}) in to yield
\begin{subequations}   \label{indyn3}
\begin{align}
\frac{\d x_{\pm}}{\d t} &= - \frac{1}{\alpha_{\pm}(t)} \left[ \int_{x_-(t)}^{x_+(t)} w_u(x_{\pm}(t)- y) \d y \right. \\
& \hspace{8mm} \left. \pm  v(t) (w_u(0) - w_u(x_+(t) - x_-(t))) - \theta_u \right], \nonumber\\
\frac{\d \Delta_{\pm}}{\d t} & = - \frac{1}{\beta_{\pm}(t)} \left[ \int_{\Delta_-(t)}^{\Delta_+(t)} w_q(\Delta_{\pm}(t),y) \d y \right. \label{indyn3b} \\
& \hspace{8mm} \left. + \int_{x_-(t)}^{x_+(t)} w_p(\Delta_{\pm}(t)-y) \d y - \theta_q \right].   \nonumber
\end{align} 
\end{subequations}
To further simplify Eq.~(\ref{indyn3}), we assume $x_+(t) - x_-(t) \approx b-a$, based on our linear stability analysis showing bumps are stable to width perturbations, but marginally stable to position perturbations. This is means we assume $u(x,t) \approx U(x - \Delta_u(t))$, so the neural activity variable is roughly a temporal translation of a stationary bump~\cite{kilpatrick13}. In this case, the first integral term in Eq.~(\ref{indyn3}) and $\theta_u$ cancel. Thus, $w_u(0) - w_u(x_+(t) - x_-(t)) \approx |U'(a)|$. We also approximate the spatial gradients using the stationary solutions, $\alpha_{\pm}(t) \approx \mp U'(a)$, $\beta_- (t) \approx Q'(c)$, and $\beta_+(t) = Q'(d)$. Such an approximation will hold in the limit of small changes to the location of the bump and front interfaces, $u(x,t) = U(x) + \ve \Psi (x,t) + {\mc O}(\ve^2)$ and $q(x,t) = Q(x) + \ve \Phi (x,t) + {\mc O}(\ve^2)$, since then $u_x(x,t) = U'(x) + {\mc O}(\ve)$ and $q_x(x,t) = Q'(x) + {\mc O}(\ve)$.  Since the terms inside the brackets of Eq.~(\ref{indyn3b}) are ${\mc O}(\ve)$, performing an ${\mc O}(1)$ approximation of $u_x$ and $q_x$ constitutes a linear approximation of the dynamics of $\Delta_{\pm}(t)$. This combination of approximations leads to a simplified set of interface equations. The edges of the bump propagate according to the velocity input
\begin{align}
x_{\pm}(t) &= \int_0^t v(s) \d s + x_{\pm}(0), \label{xint}
\end{align}
so we assign $\Delta_u(t) = (x_+(t) + x_-(t))/2$ to be the location of the bump in the position layer, and
\begin{align*}
\Delta_u(t) & =  \int_0^t v(s) \d s + \Delta_u(0), 
\end{align*}
where assuming $x_+(t) - x_-(t) \approx b-a = 2h$, $x_{\pm}(t) = \Delta_u(t) \pm h$. The front interface equations are then approximated with the stationary gradient assumptions discussed above:
\begin{align}
\frac{\d \Delta_+}{\d t} &= \gamma_d \left[ \int_{\Delta_-(t)}^{\Delta_+(t)} w_q(\Delta_+(t),y) \d y  \right. \label{fface} \\
& \hspace{15mm} \left. + \int_{\Delta_+(t) - \Delta_u(t) - h}^{\Delta_+(t) - \Delta_u(t) + h} w_p(y) \d y - \theta_q \right], \nonumber \\
\frac{\d \Delta_-}{\d t} &= - \gamma_c \left[ \int_{\Delta_-(t)}^{\Delta_+(t)} w_q(\Delta_-(t), y) \d y \right. \nonumber \\
& \hspace{15mm} \left. + \int_{\Delta_-(t) - \Delta_u(t) - h}^{\Delta_-(t) - \Delta_u(t) + h} w_p(y) \d y - \theta_q \right], \nonumber
\end{align}
where $\gamma_c^{-1} = |Q'(c)|$ and $\gamma_d^{-1} = |Q'(d)|$, so $\Delta_u(t)$ integrates the weak velocity input, and the front interfaces $\Delta_{\pm}(t)$ interact nonlinearly with the location of the bump. 

\begin{figure*}
\begin{center} \includegraphics[width=14cm]{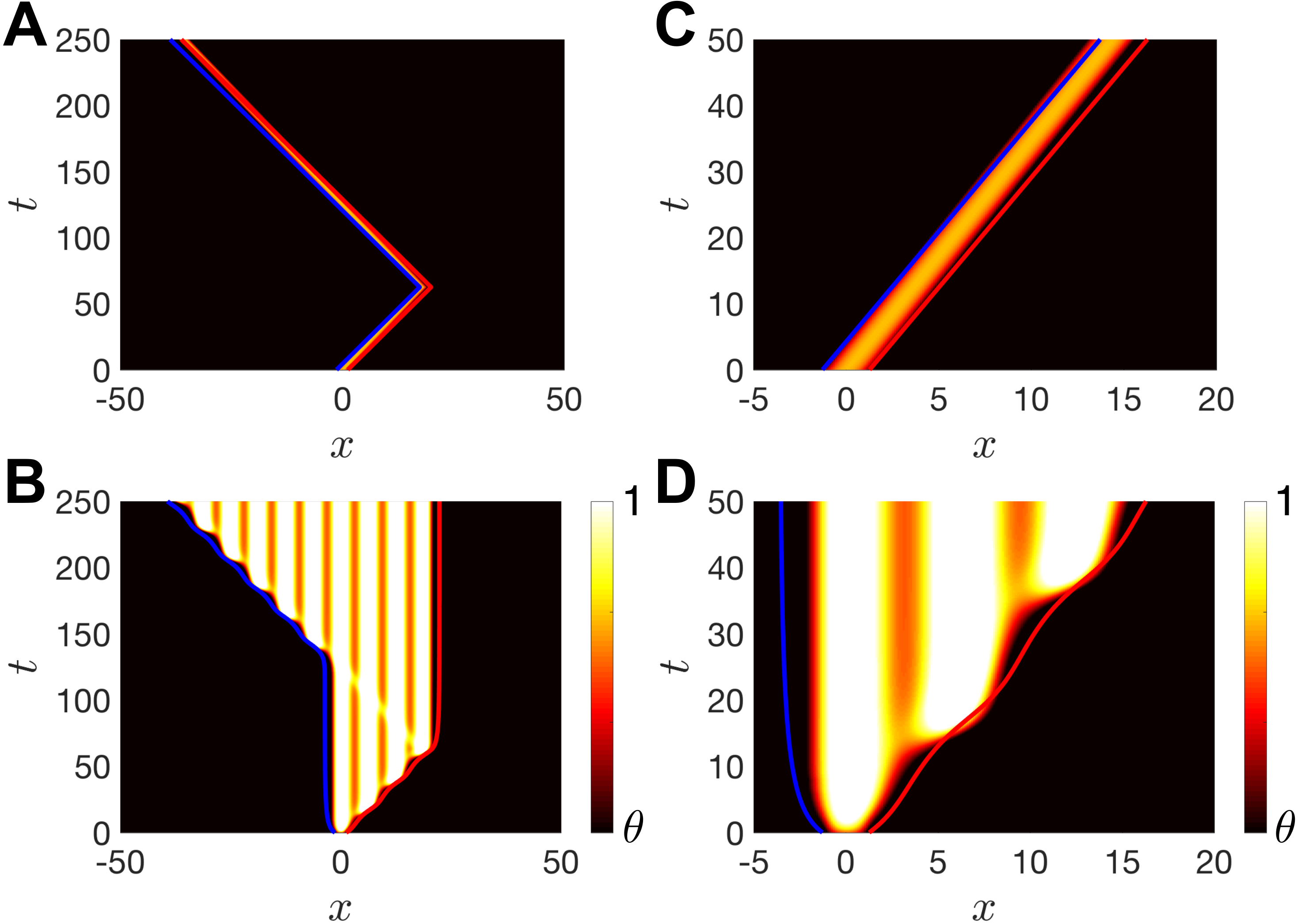} \end{center}
\caption{Interface equations approximate the dynamics of the full neural field model, Eq.~(\ref{pfield}) and (\ref{mfield}). {\bf A.} Bump propagates across the domain of the position layer, $u(x,t)$, in response to a velocity input defined $v(t) = 0.3$ on $t \in [0, 62.5)$ and $v(t) = -0.3$ on $t \in [62.5, 250]$. Interfaces defined by Eq.~(\ref{xint}) approximately track the threshold crossing locations $u(x_{\pm}(t),t) = \theta_u$ of the full simulation. {\bf B.} Memory layer, $q(x,t)$, supports a front solution that propagates in response to the motion of the bump in the position layer. Our interface approximation, $\Delta_{\pm}(t)$, given by Eq.~(\ref{fface}) correspondingly tracks the left and right boundaries of the visited regions of the searching agent. {\bf C,D.} Zoomed-in versions of the simulations in {\bf A,B}, showing slight mismatches in the approximation that occur due to our truncations. Colorbar labels show minimal color corresponds to the threshold value $\theta$ of the layer in each plot ($\theta_u$ for $u(x,t)$ and $\theta_q$ for $q(x,t)$), while 1 is the maximal color value. Parameters are $\theta_u = 0.2$, $\theta_q = 0.4$, $n=1$, $\sigma = 0.3$, $I_0 = 0.2$, and $\alpha = 1$.}
\label{fig6_lowdim}
\end{figure*}

To simplify Eq.~(\ref{fface}) further, we compute integrals, recalling $w_q(x,y)$ and $w_p(x)$ are given by Eq.~(\ref{wq}) and (\ref{wp}). The corresponding integrals are essentially the same as those evaluated in Eq.~(\ref{stateqns}) and (\ref{statsoln}). The resulting formulas simplify, since we are only examining the dynamics along the interfaces:
\begin{align}
\frac{\d \Delta_+}{\d t} &= \gamma_d \left[ \sigma \frac{\cos (n \Delta_+(t)) + n \sin (n \Delta_+(t))}{2(n^2+1)} \right.  \label{fface2} \\
& \hspace{12mm} \left. + I_0 G(\Delta_+(t) - \Delta_u(t)) +   \frac{1}{2} - \theta_q \right], \nonumber \\
\frac{\d \Delta_-}{\d t} &= - \gamma_c \left[ \sigma  \frac{\cos (n \Delta_-(t)) - n \sin (n \Delta_-(t))}{2(n^2+1)} \right. \nonumber \\
& \hspace{12mm} \left. + I_0 G(\Delta_-(t) - \Delta_u(t)) + \frac{1}{2} - \theta_q \right], \nonumber
\end{align}
where $G(\Delta) = {\mc S}(h - \Delta) + {\mc S}(\Delta+h)$ and ${\mc S}(x) = \sign (x) \left( 1 - \e^{-\alpha|x|} \right)$, and $\e^{-|\Delta_+ - \Delta_-|} \approx 0$, assuming large interface separation. Assuming $\Delta_+ - \Delta_- \gg 1$ also allows us to approximate the spatial gradients: $\gamma_c = \gamma_d$, so it suffices to use Eq.~(\ref{iface2b}) in the weak input, $I_0 \ll 1$, limit, yielding
\begin{align*}
d &= \frac{2}{n} \tan^{-1} \left(  \frac{\sqrt{\sigma^2 + n^2 \sigma^2 - (2 \theta_q-1)^2(n^2+1)^2} + n \sigma }{\sigma + (2 \theta_q-1)(n^2+1)} \right),
\end{align*}
up to periodicity, so that
\begin{align*}
\gamma_c^{-1} = \gamma_d^{-1} = - Q'(d) = \frac{1}{2} \left[ 1 + \sigma \frac{\cos (nd) + n \sin (nd)}{n^2 + 1} \right].
\end{align*}

We can now notice a number of features of the full system Eq.~(\ref{mfield}) captured by the interface Eq.~(\ref{fface2}). First, in the absence of any heterogeneity ($\sigma = 0$) or positional input ($I_0 = 0$), the front interfaces propagate at a speed approximated by $\gamma_d (1/2 - \theta_q)$ on the right ($\Delta_+(t)$) and $\gamma_c (1/2 - \theta_q)$ on the left ($\Delta_-(t)$). Sufficiently strong heterogeneity ($\sigma = \sigma^c > 0$) will pin the front. Without any positional input ($I_0 = 0$), the critical value $\sigma^c$ that pins fronts is given by the $\sigma$ such that the maximum of $\sigma \left[ \cos (n \Delta_+) + n \sin (n \Delta_+) \right]$ equals $n^2+1 - 2 \theta_q$. This occurs when $\sigma^c = \left[ n^2+1 - 2 \theta_q \right] / \left[ \cos {\mc T}(n) - n \sin {\mc T}(n)  \right]$ for ${\mc T}(n) = 2 \tan^{-1}((1-\sqrt{n^2+1})/n)$, corresponding to the critical heterogeneity for wave propagation failure discussed in \cite{bressloff01,coombes11}. Thus, we require $\sigma > \sigma^c$ for the system to retain memory of visited locations, which prevents front propagation to the rest of the domain.

Our interface equations are compared with simulations of the full model Eq.~(\ref{pfield}) and (\ref{mfield}) in Fig. \ref{fig6_lowdim}. The evolution of the bump interfaces in the positional layer $u(x,t)$ ($u(x_{\pm}(t), t) = \theta_u$) are captured well by $x_{\pm}(t) = \Delta_u(t) \pm h$ (Fig. \ref{fig6_lowdim}A,C). We expect the mismatch arises as the result of our static gradient approximation $u_x(x_{\pm}(t),t) \approx \pm U'( \pm h)$. The front tracks previously visited locations of the bump, corresponding to the active regions in the domain at time $t$ (Fig. \ref{fig6_lowdim}B,D). More regions are activated when the searcher position enters an unvisited part of the domain. Otherwise, the front solution remains stationary. Thus far, we have utilized an open-loop velocity protocol, so that the velocity input to the position layer does not receive feedback from the memory layer.

Our low-dimensional approximation, Eq.~(\ref{fface}), performs well when compared with numerical simulations. Thus, we have established a mechanism by which a ballistic searcher may store a memory of previously visited locations. In the next section, we analyze our reduced model, and study search strategies whereby an agent's behavior depends on which regions it has searched.

\section{Memory-guided search}
\label{perform}

Now, we study the impact of memory-guided control on the efficiency of a searching agent with memory for previously visited locations.
First, we consider search along a single segment, and control speeds up search when the agent arrives at a previously visited location. Assuming finding of a target is stochastic, this reduces time spent in already visited locations. Interestingly, this does not appear to reduce search time, and the optimal search speed is the same whether in novel or previous searched territory. Second, we consider a radial arm maze, changing the geometry from a single segment to multiple segments connected at one end. An IOR strategy is advantageous in this case, since it prevents the agent from searching previously searched arms, in the initial exploratory phase of the search. Our theory is then compared to Monte Carlo simulations of ballistic searcher model, which we describe below. \\
\vspace{-3mm} 

\noindent
{\bf Single segment.} We begin by exploring memory-guided search in a single segment. Avoidance of previously visited locations will not work in this case, since the searcher would be trapped at the end of the segment, so we invoke a strategy whereby the searcher changes its speed when exploring a previously visited portion of the segment. We map the activities of the position $u(x,t)$ and memory ($q(x,t)$) layers directly onto a velocity variable:
\begin{align}
v(t) = \chi (u(x,t),q(x,t)) \cdot v_o(t),  \label{vcontrol}
\end{align}
so that $v_o(t)$ is an open-loop component of the velocity control, not subject to the internal neural activity variables. The open-loop component is then modulated by a closed-loop control $\chi(u,q)$ with evolution equation
\begin{align}
\tau_{\chi} \dot{\chi}(t) =& 2\langle H(u - \theta_u), H(q - \theta_q) \rangle (\chi_{+} - \chi (t)) \nonumber
\\
& \hspace{11mm}
-  \langle H(u - \theta_u), 1 \rangle (\chi_{-} - \chi(t)),  \label{chi}
\end{align}
so when the position layer totally overlaps with the memory layer, both inner products equal $\Delta_u(t)$, the width of the bump. The steady state control input in this case of total overlap is $\chi (t) \to 2 \chi_+ - \chi_-$, whereas when the position layer does not overlap with the memory layer, $\chi (t) \to \chi_-$.
Search will speed up (slow down) when the agent is in a location it has already visited for $\chi_+ > \chi_-$ ($\chi_-/2 < \chi_+ < \chi_-$). In this way, memory of previous positions can guide search.


\begin{figure*}
\begin{center} \includegraphics[width=17cm]{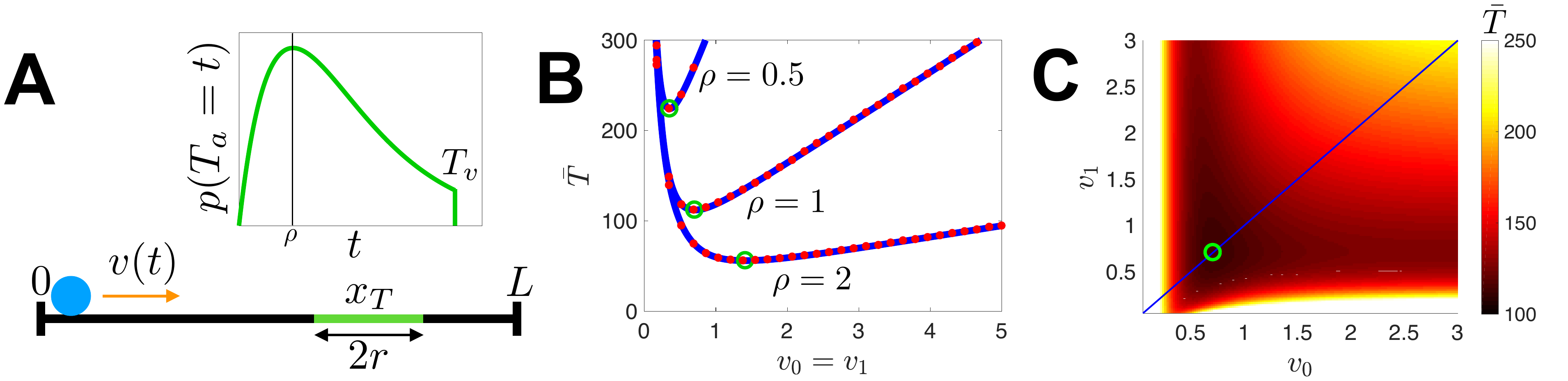} \end{center}
\caption{Ballistically moving agent searches for a hidden target. {\bf A}. Searcher (dot) begins at the left edge ($x=0$) of the domain ($x \in [0,L]$), initially moving with speed $v_0$ and then moving with speed $v_1$ on all subsequent trips across. The target (green line) spanning $x\in [x_T-r,x_T+r]$ is stochastically discoverable according to the waiting time density $p(t) = \rho^2 t\e^{- \rho t}$ (plot above), so if the waiting time exceeds $T_v$, the searcher will not find the target on the current trip. {\bf B}. Plots of $\bar{T}$ versus $v_0 = v_1$ (line) using Eq.~(\ref{Tv0}) are nonmonotonic, revealing an interior optimum that minimizes the average search time (circles). As the rate of target discovery $\rho$ decreases, $\bar{T}$ increases, and the optimal $v_0$ decreases. Theory matches well with averages from $10^6$ Monte Carlo simulations (dots). {\bf C}. Mean search time $\bar{T}$ as a function of both $v_0$ and $v_1$, showing the optimal choice $(v_0,v_1)$ occurs when $v_0 = v_1 \approx 0.706$ (circle) when $\rho = 1$. Other parameters are $L=100$ and $r=1$.}
\label{fig7_ballistic}
\end{figure*}

Now, to analyze the impact of this memory-guided search strategy, we consider the limit of the interface Eq.~(\ref{fface2}) in which a searcher starts in the left side of a bounded domain. All our calculations to this point assumed $x\in (- \infty, \infty)$, an approximation to the case of large bounded domains of length $L \gg 1$ we analyze now. In this case, the left interface $\Delta_-(t)$ is irrelevant. Furthermore, we study the limit in which the position input dominates the dynamics. Taking $\theta_q \to 1/2$, $\sigma \to 0$ and assuming a velocity with constant amplitude to start $|v(t)| = v_0$, we then have
\begin{align*}
\frac{\d \Delta_+}{\d t} = \gamma_d I_0 G( \Delta_+(t) - v_0 t)
\end{align*}
by plugging into Eq.~(\ref{fface2}), where we have $|\Delta_u(t)| = v_0 t$. Switches in the direction of the velocity occur when the agent encounters the boundary of the domain, $v(t) \mapsto - v(t)$. The agent begins moving rightward: $v(t) = v_0 > 0$. The corresponding phase-shift between the front interface position and the bump position is $\Omega_0 = v_0t - \Delta_+(t)$, so $\Delta_+(t) = v_0t - \Omega_0$. We thus expect constant velocity solutions for $\Delta_+(t)$ as long as the condition
\begin{align*}
v_0 = \gamma_d I_0 G( - \Omega_0),
\end{align*}
holds for some phase-shift $\Omega_0$. Note also since we are using closed loop control, this solution will only be valid if $v_0$ is the constant velocity to which the control loop has equilibrated. Self-consistency of Eq.~(\ref{vcontrol}) and (\ref{chi}) then require $v_0 = \bar{\chi}_0 \cdot v_o$, where
\begin{align*}
\bar{\chi}_0 = \frac{2{\rm max}(h- \Omega_0,0) \chi_+ - 2 h \chi_-}{2 {\rm max}(h- \Omega_0,0) - 2 h},
\end{align*}
where $h$ is the half-width of the bump in the position layer. Once the domain has been searched, the control variable is updated to $\bar{\chi}_1 = 2\chi_+ - \chi_-$, and we thus define $v_1= \bar{\chi}_1 \cdot v_o$. Since the parameters $\chi_{\pm}$ can be tuned to give any pair $v_{0,1}$, we focus on the limit $\tau_{\chi} \to 0$, and assume the agent employs these two search velocities, depending on a novel or searched regime.

We model the agent's search behavior as follows. The agent enters a bounded interval $x \in [0,L]$ from the left side ($x=0$), and searches ballistically at a constant speed, determined by whether it is in the unsearched ($v_0$) or searched domain ($v_1$). The finite-sized target has radius $r$ and is centered at $x_T$, so it spans $x \in [x_T - r, x_T +r]$ and $r \leq x_T \leq L - r$ (Fig. \ref{fig7_ballistic}A). The agent discovers the target stochastically, according to a Gamma distributed waiting time, $\rho^2 \e^{-\rho t}$. Such dynamics could arise for a two-stage process whereby the searcher first realizes an object of interest is nearby, and then compares it with the target object from memory. For velocity $v$, the agent is over the target for a time $T_v = 2r/v$, so the probability of discovering the target on a single trip over is
\begin{align}
P_v = \rho^2 \int_0^{T_v} t \e^{- \rho t} \d t = 1 - (1 + \rho T_v) \e^{- \rho T_v},  \label{Pv}
\end{align}
and the associated conditional mean time to find the target while over it is
\begin{align}
T_a(v) &=  \frac{\rho^2}{P_v} \int_0^{T_v} t^2 \e^{- \rho t} \d t \nonumber \\
&= \frac{1}{\rho P_v} \left[ 2 - (2 + 2 \rho T_v + \rho^2 T_v^2) \e^{- \rho T_v} \right].   \label{Ta}
\end{align}

We now address the problem of finding the velocities $(v_0,v_1)$, corresponding to the novel and searched territory, that minimize the time to find the target. The mean first passage time can be derived analytically by tracking the probability of absorption and accumulated search time at each target encounter. The first visit to the target occurs after $T_L(v_0) = (x_T-r)/v_0$. During the first pass over the target, the searcher discovers the target with probability $P_{v_0}$, Eq.~(\ref{Pv}), with conditional mean time within the target $T_a(v_0)$, Eq.~(\ref{Ta}). The time between the first and the second visits is $T_R(v_0)+T_R(v_1)$, where $T_R(v) = (L-x_T-r)/v$, and the probability of finding the trap during the next visit is $P_{v_1}$ with mean time $T_a(v_1)$. Subsequent times and probabilities are computed similarly, and the time spent searching scales linearly with the length of the searcher's path. Using geometric series, we can compute the mean time to find the target by marginalizing over all possible visit counts
\begin{align}
&T (x_T)  = T_L(v_0) + P_{v_0}  T_a(v_0) + \frac{1- P_{v_0}}{2-P_{v_1}} \left[ \frac{2L}{v_1 P_{v_1}} \right. \nonumber \\
& \left. + (1- P_{v_1}) \left( T_{v_0} + T_R(v_0) + T_L(v_1) + T_a(v_1) - \frac{L}{v_1} \right) \right. \nonumber\\
& \left. + T_{v_0} + T_R(v_0) + T_R(v_1) + T_a(v_1) - \frac{2L}{v_1} \right].  \label{TxT}
\end{align}
The generalized mean first passage time is then given by integrating over the range of possible target locations $x_T$, assuming a uniform probability of placement: $\bar{T} = \frac{1}{L-2r} \int_{r}^{L-r} \langle T (x_T) \rangle \d x_T$. Since the only terms in Eq.~(\ref{TxT}) that depend on $x_T$ are $T_L(v)$ and $T_R(v)$, we need only compute $\bar{T}_L(v) = \frac{1}{L - 2r} \int_r^{L - r} \frac{x_T - r}{v} \d x_T = \frac{L - 2r}{2v}$ and $\bar{T}_R(v) = \frac{1}{L - 2r} \int_r^{L - r} \frac{L-x_T - r}{v} \d x_T = \frac{L-2r}{2v}$,
and we rescale space, so it is in units of the radius $r$. This is equivalent to setting $r=1$ in Eq.~(\ref{TxT}), and considering any spatial parameters as in rescaled coordinates, which yields
\begin{align}
\bar{T}  =& \frac{L-2}{2 v_0} + P_{v_0}  T_a(v_0) + (1- P_{v_0}) \left[ \frac{L}{v_1 P_{v_1}} \right. \nonumber \\
& \left. \hspace{12mm} + \left( 1 + \frac{L}{2} \right) \left( \frac{1}{v_0} - \frac{1}{v_1} \right) + T_a(v_1) \right].  \label{gmfpt}
\end{align}
Note, for constant speeds $v_1 = v_0$, Eq.~(\ref{gmfpt}) simplifies considerably to
\begin{align}
\bar{T} (v_1 \equiv v_0) = \frac{L}{2 P_{v_0} v_0} (2- P_{v_0}) + T_a(v_0) - \frac{1}{v_0}.   \label{Tv0}
\end{align}
As shown in Fig. \ref{fig7_ballistic}B, $\bar{T}(v_1 = v_0)$ has an internal minimum, which leads to the most rapid finding of the target. Notably, in Fig. \ref{fig7_ballistic}C, we find there is no advantage in searching more quickly (or slowly), once the domain has already been searched. In fact, the search time is minimized when $v_1 = v_0$.

Thus, for single segments, memory-guidance does not speed up search, in this particular paradigm. The optimal strategy for minimizing the time to find the target is for the searcher to maintain the same search speed throughout the exploration process. We now demonstrate an alternative paradigm in which memory-guided search does reduce the time to find the target. \\
\vspace{-3mm}

\begin{figure*}
\begin{center} \includegraphics[width=17cm]{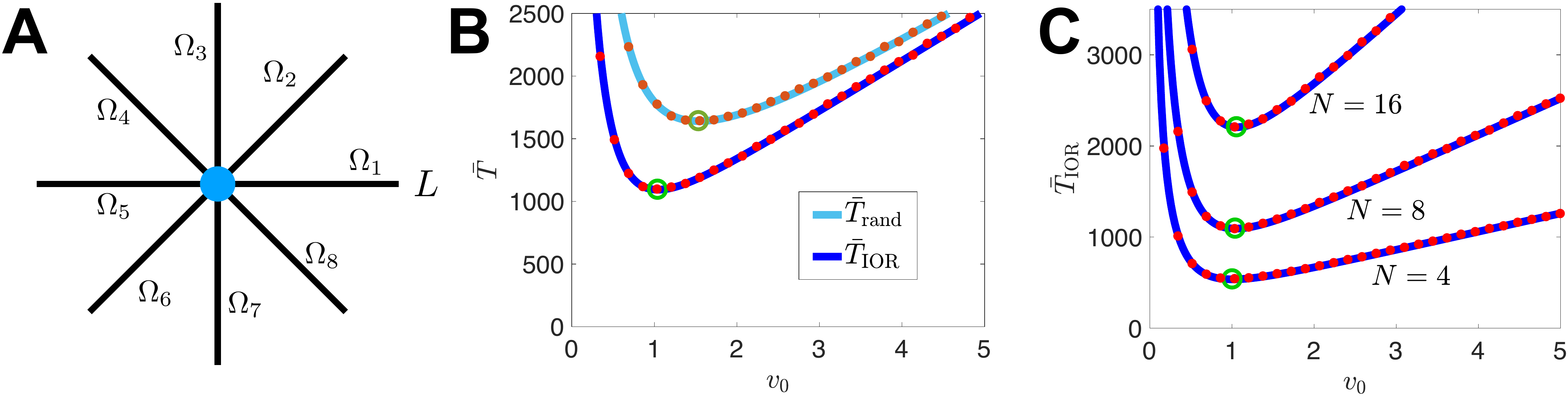} \end{center}
\vspace{-4mm}
\caption{Ballistically-moving agent searches a radial arm maze with a single target in a single arm. {\bf A}. Searcher (dot) begins at the center of the maze, and chooses a random arm $\Omega_k$, $k \in \{1,...,N\}$ to search. Purely random search (rand) proceeds with the searcher always choosing $1$ out of $N$ total arm. Inhibition-of-return (IOR) guides searcher away from previously searched arms, so the first $N$ arms chosen are the arms $k=1,...,N$. {\bf B}. Inhibition-of-return leads to more rapid location of the target than purely random search ($\bar{T}_{\text{rand}} > \bar{T}_{\text{IOR}}$ as in Eq.~(\ref{branchineq}). Theory (solid lines) matches $10^6$ Monte Carlo simulations (dots) very well. Here $N=8$. {\bf C}. Average time to find the target $\bar{T}_{\text{IOR}}$ using IOR increases with the number of arms, but note the optimal search speed $v_0^{\text{min}}$ (circles) remains relatively unchanged. Other parameters are $L=100$, $r=1$, and $\rho = 1$.}
\label{fig8_ram}
\end{figure*}

\noindent
{\bf Radial arm maze.} Since search on a single segment does not appear to be aided my memory-guidance, we examine the case in which the agent must search over a space with more complex topology. In particular, we study the problem of the searcher finding a hidden target in a radial arm maze (Fig. \ref{fig8_ram}A). This paradigm has commonly been used to test mammalian memory, requiring a combination of spatial navigation, decision-making, and working memory~\cite{olton77,floresco97}. Rather than deriving a new neural field model and associated interface equations on this more complex domain, we develop a simpler model for memory of previously visited locations using a metastable neuronal network with distinct populations encoding each arm. We assume the searcher begins at the center of the maze with $N$ arms that radiate outwards, so locations lie on the union of bounded intervals $\Omega_1 \cup \Omega_2 \cup \cdots \cup \Omega_N$ with $\Omega_j = [0,L]$ for all $j$. The target lies within one of the arms $k \in \{1, ..., N\}$ at a location $x_T \in [r,L-r]$ as before. Since our previous analysis did not reveal an advantage to storing the spatial structure of locations visited within a segment, we remark that memory of visited arms can be stored by distinct bistable neural populations:
\begin{align}
\dot{q}_j(t) &= -q_j(t) + H(q_j - \theta_q) + I_j(t),  \label{ramq}
\end{align}
where $I_j(t) = I_0> \theta_q$ when the agent visits arm $j$ and $I_j(t) =0$ otherwise. The variables $q_j(t) \to 1$ once arm $j$ is visited, and initially $q_j(0) = 0$ for all $j$. If the searcher avoids arms such that $q_j(t) > \theta_q$, they will only visit novel arms until $q_j(t) \to 1$ for all $j$. Thus, Eq.~(\ref{ramq}) constitutes a discretized version of Eq.~(\ref{mfield}). When the searcher is over the target, it discovers it according to a Gamma distributed waiting time. The probability of discovering the target at each encounter is $P_v$, Eq.~(\ref{Pv}), and the conditional mean first passage time within the target is $T_a(v)$, Eq.~(\ref{Ta}).

We now derive the mean time to find the target, as in the case of a single armed domain. In particular, we compare the effects of IOR, where the searcher avoids previously explored arms initially, as opposed to a memoryless selection of the next arm to be searched. As mentioned, we assume the speed of search is constant throughout the process $|v(t)| \equiv v_0$ for all $t$. Following the steps of our previous calculation (and see also \cite{poll16}), we find that (for $r=1$) the average time for a memoryless to find a target placed uniformly on $x_T \in [1,L-1]$ on one of $N$ radial arms is
\begin{align}
\bar{T}_{\text{rand}}  =& \frac{2L (N-1)}{v_0} + \frac{2NL (1-P_{v_0})^2}{P_{v_0}(2-P_{v_0}) v_0} \label{ramrand} \\
& \hspace{12mm} + \frac{L(1-P_{v_0})}{(2-P_{v_0})v_0} + \frac{L-2}{2v_0} + T_a(v_0). \nonumber
\end{align}
On the other hand, a searcher that uses IOR to avoid previously explored arms prior to all arms being searched finds the target after an average time
\begin{align}
\bar{T}_{\text{IOR}}  =& \frac{L (N-1)}{v_0} + \frac{2NL (1-P_{v_0})^2}{P_{v_0}(2-P_{v_0}) v_0} \label{ramrand} \\
& \hspace{12mm} + \frac{L(1-P_{v_0})}{(2-P_{v_0})v_0} + \frac{L-2}{2v_0} + T_a(v_0),\nonumber
\end{align}
which appears nearly the same as the random search time $\bar{T}_{\text{rand}}$, except that the leading factor is roughly half for $\bar{T}_{\text{IOR}}$. In fact, it is straightforward to show $\bar{T}_{\text{rand}} \geq \bar{T}_{\text{IOR}}$ for any $N \geq 2$, since
\begin{align}
\bar{T}_{\text{rand}} - \bar{T}_{\text{IOR}}  = \frac{N-1}{2} \frac{L}{v_0} > 0,  \label{branchineq}
\end{align}
for $N \geq 2$. This theory is matched very well to Monte Carlo simulations of the ballistic searcher (Fig. \ref{fig8_ram}B), demonstrating the efficacy of IOR in reducing the time to find the target. This effect is even stronger for mazes with more arms (higher $N$) as the total time to find the target (Fig. \ref{fig8_ram}C) and the discrepancy between IOR and random search increases with $N$.

\begin{figure*}
\begin{center} \includegraphics[width=17cm]{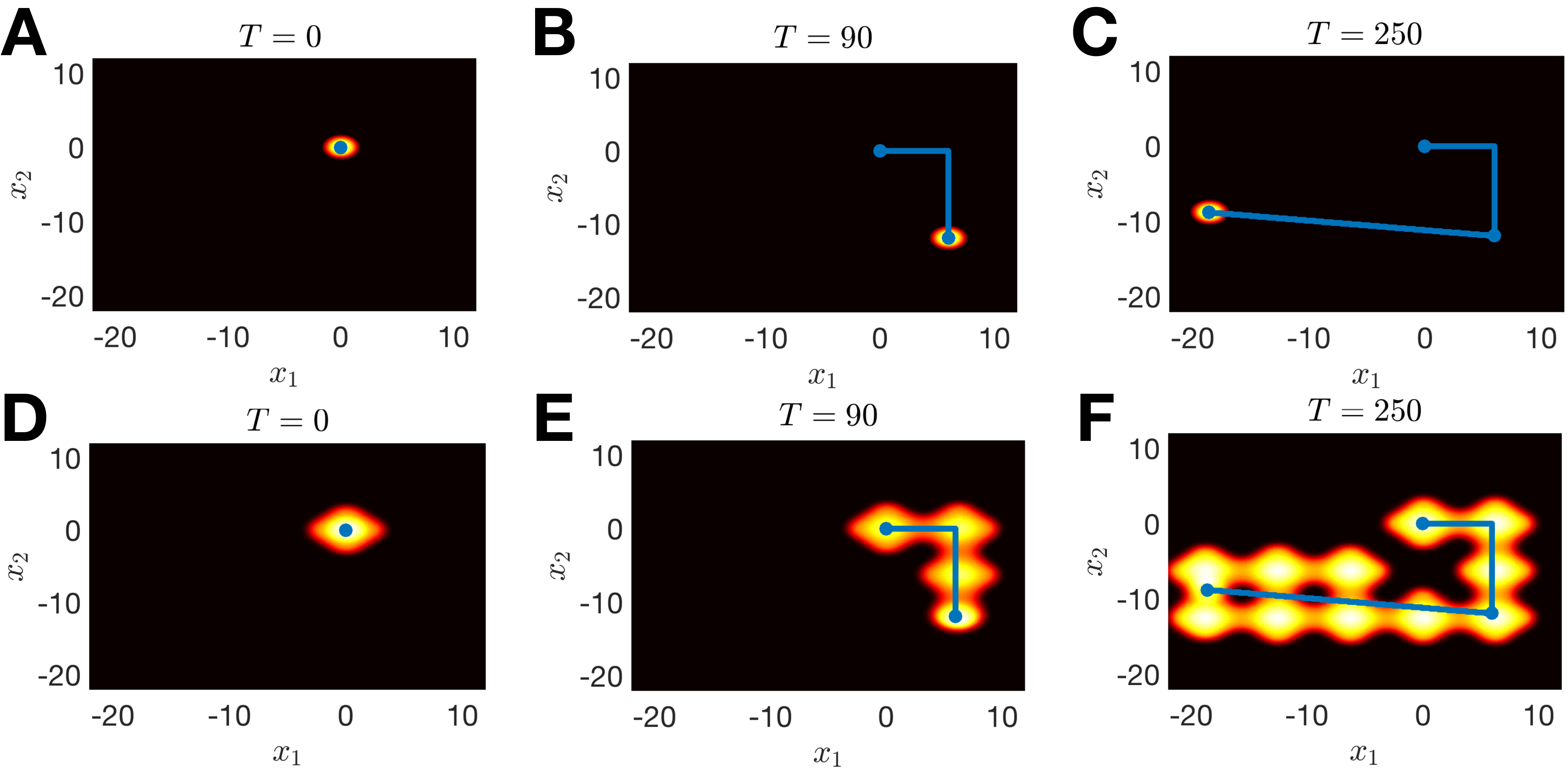} \end{center}
\caption{Two-dimensional simulation of the neural field model, Eq.~(\ref{2dfield}), describing the propagation of a coupled bump and front in a planar domain. Evolution of the bump in the position layer $u(\x,t)$ is tracked by showing snapshots at ({\bf A}) $T=0$; ({\bf B}) $T=90$; and ({\bf C}) $T=250$. The path of the agent is shown by the solid blue line. Motion of the bump layer is stored by the front layer $q(\x,t)$, which tracks the previously visited locations of the bump at the same snapshots in time: ({\bf A}) $T=0$; ({\bf B}) $T=90$; and ({\bf C}) $T=250$. Parameters are $\theta_u = 0.2 $, $\theta_q = 0.45$, $n=1$, $\sigma = 0.3$, $\upsilon = 0.3$, and $I_0 = 0.3$.}
\label{fig9_2dsim}
\end{figure*}

Our analysis of the neural field model has demonstrated a plausible neural mechanism for memory-guided search, persistent activity encoding previously searched regions. The theory and simulations we have performed here for corresponding ballistic searcher models has demonstrated that memory-guided search does not appear to be advantageous in one-dimensional domains comprised of a single segment. However, multiple segments adjoined at there ends can comprise more complex domains like the radial arm maze, which do benefit from inhibition-of-return (IOR). A searcher that avoids previously searched segments will tend to find a randomly placed target more quickly than a searcher that chooses subsequent arms in a memoryless way. Since our low-dimensional theory was derived from the full neural field equations, we expect that stochastic simulations of the full neural field model would yield qualitatively similar results. We now discuss briefly how our theory might be extended to two-dimensional domains.

\section{Extensions to two-dimensions}
\label{2D}

Most visual and navigational search tasks tend to be in spaces of two or more dimensions (See \ref{fig1_examples}D,E and \cite{hills15,mueller08,kanitscheider17}). In future work, we will extend our analysis of our one-dimensional model, Eq.~(\ref{pfield}) and (\ref{mfield}), to an analogous two-dimensional model. In this case, we expect there to be a wider variety of control mechanisms that lead to an efficient use of memory in guiding future search locations. Analysis of stationary solutions in two-dimensional neural field models has been successful in a number of cases~\cite{laing03,folias04,owen07,gokcce17}, and there is a clear path to extending interface methods to describe contour boundaries that arise for solutions in planar systems~\cite{coombes12}.

Here we briefly discuss a candidate model for memory-guided search in two-dimensions. In particular, we will demonstrate in numerical simulations that such a model does result in a model that can store previously visited locations in the plane. Memory of a searching agent's position and memory for previously searched locations are captured by the following pair of neural field equations on a planar domain:
\begin{align}
u_t &= -u +w_u*H(u-\theta_u)- \v(t) \cdot \left(\nabla w_u \right)*H(u-\theta_u), \nonumber \\
q_t &= -q +w_q*H(q-\theta_q) + w_p*H(u- \theta_u), \label{2dfield}
\end{align}
defined on $\x = (x_1, x_2)^T \in \R^2$. Recurrent coupling in the position layer is described by the integral $w_u*H(u - \theta_u) = \int_{ \R^2} w_u( \x - \y) H(u(\y,t) - \theta_u) \d \y $, and the synaptic kernel is lateral inhibitory and rotationally symmetric ($w_u(\x,\y) = w_u(z)$, $z = \sqrt{(x_1-y_1)^2+(x_2-y_2)^2}$) comprised of a difference of Bessel functions of the second kind~\cite{owen07}:
\begin{align*}
w_u(z) = \sum_{k=1}^4 c_k K_0(\alpha_k z),
\end{align*}
with $[c_1, c_2, c_3, c_4] = [5/3,-5/3,-1/2,1/2]$ and $[\alpha_1,\alpha_2,\alpha_3,\alpha_4] = [1,2,1/4,1/2]$. 
Velocity input is given by a two-dimensional vector $\v (t) = (v_1(t),v_2(t))^T$, which translate bumps when taking its dot product with the gradient of the weight function $\nabla w_u(r) = (\partial_{x_1}, \partial_{x_2})^T w_u(r)$, $r =  \sqrt{x_1^2+x_2^2}$.
The heterogeneous connectivity function that pins the activity in the memory layer is defined using the product of cosines and an exponential:
\begin{align}
w_q(\x,\y) = (1 + \sigma \cos (n y_1) + \sigma_2  \cos (n y_2)) \frac{\e^{-\upsilon d^2}}{2 \pi}.  \label{hetwt2d}
\end{align}
As in the one-dimensional case, the weight function is a homogeneous kernel modulated by periodic heterogeneities. We will demonstrate in numerical simulations that these heterogeneities can pin the expansion of wave fronts, analogous to the stabilizing effects they have on stationary bumps in planar neural fields~\cite{poll15}. Lastly, we consider an input term from the position layer, applying feedforward input centered at the location of the bump:
\begin{align*}
w_p(d) =  I_0 \e^{-\upsilon z^2}.
\end{align*}

We now demonstrate that this model is capable of generating a memory trace for previously visited regions of a searcher exploring two-dimensional space. In Fig. \ref{fig9_2dsim}, we demonstrate the results from a numerical simulation of the neural field Eq.~(\ref{2dfield}). A bump is instantiated in the position layer $u(\x,t)$, and tracks the locations visited by an agent moving about the domain (Fig. \ref{fig9_2dsim}A-C), evolving in response to velocity inputs. The motion of the bump is reflected by the memory of previously visited locations tracked by the front layer $q(\x,t)$ (Fig. \ref{fig9_2dsim}D-F). The activity in the front layer is stabilized by the heterogeneity in the weight kernel, Eq.~(\ref{hetwt2d}), as it was in the one-dimensional case.

\begin{figure}
\begin{center} \includegraphics[width=8.3cm]{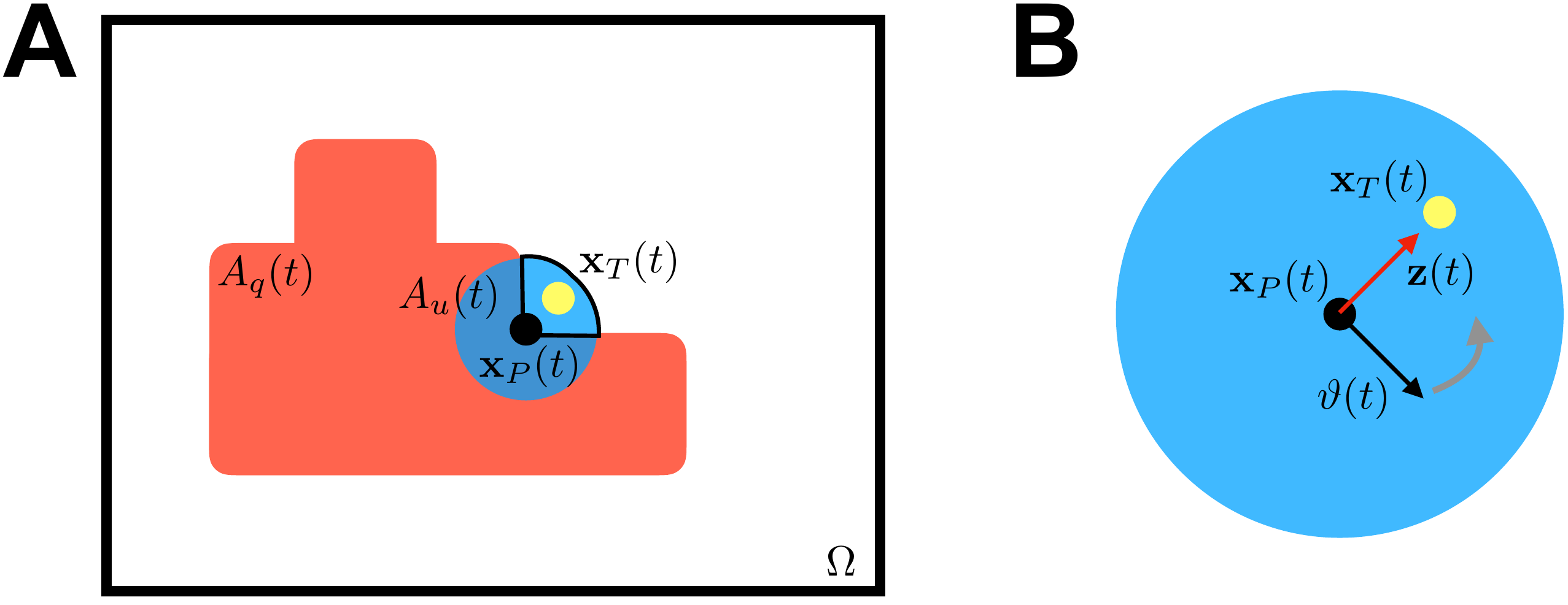} \end{center}
\caption{Inhibition-of-return is implemented when an agent moves in the direction of the unexplored regions of the domain $\Omega$. {\bf A.} The intersection of the bump layer's active region $A_u(t)$ and the complement of the active region $A_q(t)$ is the region yet to be searched (quarter section of circle) with center-of-mass $x_T(t)$ close to the agent's current location $\x_P(t)$ (black dot). {\bf B.} The agent reorients its velocity angle $\vartheta (t)$ in the direction of the unexplored region ($\x_T(t)$).}
\label{fig10_orient}
\end{figure}

We now discuss a control mechanism that we conjecture could lead to successful inhibition-of-return of a searcher with position and memory layer activity described by Eq.~(\ref{2dfield}). In particular, the position and memory layers will have active regions $A_u(t)$ and $A_q(t)$ describing the area of superthreshold within each (Fig. \ref{fig10_orient}A). From the active region $A_u(t)$, the position center-of-mass ${\bf x}_P(t)$ can be computed as the first moment. Second, the region $A_u(t) \cap (\Omega  \backslash A_q(t))$ describes where the position layer's activity intersects with the complement of the memory layer's activity, corresponding to unexplored space. If we call the center-of-mass of this region ${\bf x}_T(t)$, then an IOR mechanism might work by having the searcher move toward ${\bf x}_T(t)$. Thus, the angle of the searcher's velocity $\vartheta(t)$ should constantly orient in the direction of ${\bf x}_T(t)$ (Fig. \ref{fig10_orient}B).

Note, the computations involved in determining a gradient-descent type orientation of the searcher require some linear readouts~\cite{deneve99}, divisive normalization~\cite{olsen10,beck11}, and potential nonlinearities. Motor control circuits are capable of producing outputs that correspond to a wide range of nonlinearities, for example, built on summations of various nonlinear basis functions~\cite{churchland12}. Thus, we expect the computation we have outlined above could be implemented as a closed-loop feedback from the memory system onto a corresponding motor control circuit, but we do not propose a specific neural architecture for doing so at this time.

\section{Discussion}

We have demonstrated that a neural field model can store previously visited locations in a search task with persistent activity. In a one-dimensional model, feedforward connectivity from a continuous attractor network that encodes position can initiate memory-encoding activity in the form of stationary fronts whose spatial resolution is determined by the frequency of underlying synaptic heterogeneity (as in \cite{avitabile15}). Analysis of stationary solutions demonstrate the mechanism by which persistent activity expands in the memory layer is via a hysteresis. For strong enough input from the position layer, front positions in the vicinity of the position input undergo a bifurcation, leading to a rapid transition of the front to an adjacent stable location. We can capture the dynamics of these two layers by a low-dimensional approximation that tracks the interfaces of the front in the memory layer, and the position of the bump attractor in the position layer. This low-dimensional model is leveraged to test the impact of memory-guided search. We find that search along a single one-dimensional segment is not aided by memory-guided search, but search in more complex domains with distinct conjoined segments are. We expect that our approach can be extended to two-dimensional search processes, where memory-guided search is likely to be advantageous in most situations.

Our work contributes a new application of interface methods to neural field equations. Recently, the dynamics of labyrinthine and spot patterns in two-dimensions have been captured by the low-dimensional projection of their interfaces~\cite{coombes12,gokcce17}. This method has two advantages. First, it can lead to numerical simulation schemes that are an order of magnitude faster than simulating the full system, since the dimensionality of the problem can sometime be reduced by one. Second, it often leads to systems that are analytically tractable, allowing for a systematic study of both linear and nonlinear dynamics in the vicinity of equilibria. We leveraged both of these advantages in our work, since we were able to gain insight concerning the mechanism by which the memory layer dynamics evolved. Our work has shown that a searcher engaging memory from a neural field model storing previous positions can be approximately mapped to a ballistically-searching agent with velocity evolving according to a memory-shaped jump process. This provides a new and interesting link between neural fields, and low-dimensional models of stochastic search~\cite{benichou05,newby10}.

Biophysical models of search tend to use memoryless agents, particularly because this can make for straightforward analysis and explicit results for quantities such as the mean first passage time to find the target~\cite{viswanathan96}. However, there is evidence suggesting organisms employ memory of searched locations to find hidden targets in both foraging~\cite{simpson04} and visual search~\cite{bays12} tasks. For this reason, we think it is worthwhile to develop techniques for understanding search models in the form of stochastic processes with various forms of memory. Our study is a first step in the direction of both implementing memory-driven stochastic search process in combination with a proposed neural mechanism for their implementation.

\bibliographystyle{apsrev}
\bibliography{hybrid}

\begin{thebibliography}{66}
\expandafter\ifx\csname natexlab\endcsname\relax\def\natexlab#1{#1}\fi
\expandafter\ifx\csname bibnamefont\endcsname\relax
  \def\bibnamefont#1{#1}\fi
\expandafter\ifx\csname bibfnamefont\endcsname\relax
  \def\bibfnamefont#1{#1}\fi
\expandafter\ifx\csname citenamefont\endcsname\relax
  \def\citenamefont#1{#1}\fi
\expandafter\ifx\csname url\endcsname\relax
  \def\url#1{\texttt{#1}}\fi
\expandafter\ifx\csname urlprefix\endcsname\relax\def\urlprefix{URL }\fi
\providecommand{\bibinfo}[2]{#2}
\providecommand{\eprint}[2][]{\url{#2}}

\bibitem[{\citenamefont{Hills et~al.}(2015)\citenamefont{Hills, Todd, Lazer,
  Redish, Couzin, Group et~al.}}]{hills15}
\bibinfo{author}{\bibfnamefont{T.~T.} \bibnamefont{Hills}},
  \bibinfo{author}{\bibfnamefont{P.~M.} \bibnamefont{Todd}},
  \bibinfo{author}{\bibfnamefont{D.}~\bibnamefont{Lazer}},
  \bibinfo{author}{\bibfnamefont{A.~D.} \bibnamefont{Redish}},
  \bibinfo{author}{\bibfnamefont{I.~D.} \bibnamefont{Couzin}},
  \bibinfo{author}{\bibfnamefont{C.~S.~R.} \bibnamefont{Group}},
  \bibnamefont{et~al.}, \bibinfo{journal}{Trends in cognitive sciences}
  \textbf{\bibinfo{volume}{19}}, \bibinfo{pages}{46} (\bibinfo{year}{2015}).

\bibitem[{\citenamefont{Mueller and Fagan}(2008)}]{mueller08}
\bibinfo{author}{\bibfnamefont{T.}~\bibnamefont{Mueller}} \bibnamefont{and}
  \bibinfo{author}{\bibfnamefont{W.~F.} \bibnamefont{Fagan}},
  \bibinfo{journal}{Oikos} \textbf{\bibinfo{volume}{117}}, \bibinfo{pages}{654}
  (\bibinfo{year}{2008}).

\bibitem[{\citenamefont{Kanitscheider and Fiete}(2017)}]{kanitscheider17}
\bibinfo{author}{\bibfnamefont{I.}~\bibnamefont{Kanitscheider}}
  \bibnamefont{and} \bibinfo{author}{\bibfnamefont{I.}~\bibnamefont{Fiete}},
  \bibinfo{journal}{Current Opinion in Systems Biology}
  (\bibinfo{year}{2017}).

\bibitem[{\citenamefont{Daw et~al.}(2006)\citenamefont{Daw, O'doherty, Dayan,
  Seymour, and Dolan}}]{daw06}
\bibinfo{author}{\bibfnamefont{N.~D.} \bibnamefont{Daw}},
  \bibinfo{author}{\bibfnamefont{J.~P.} \bibnamefont{O'doherty}},
  \bibinfo{author}{\bibfnamefont{P.}~\bibnamefont{Dayan}},
  \bibinfo{author}{\bibfnamefont{B.}~\bibnamefont{Seymour}}, \bibnamefont{and}
  \bibinfo{author}{\bibfnamefont{R.~J.} \bibnamefont{Dolan}},
  \bibinfo{journal}{Nature} \textbf{\bibinfo{volume}{441}},
  \bibinfo{pages}{876} (\bibinfo{year}{2006}).

\bibitem[{\citenamefont{Cohen et~al.}(2007)\citenamefont{Cohen, McClure, and
  Angela}}]{cohen07}
\bibinfo{author}{\bibfnamefont{J.~D.} \bibnamefont{Cohen}},
  \bibinfo{author}{\bibfnamefont{S.~M.} \bibnamefont{McClure}},
  \bibnamefont{and} \bibinfo{author}{\bibfnamefont{J.~Y.}
  \bibnamefont{Angela}}, \bibinfo{journal}{Philosophical Transactions of the
  Royal Society of London B: Biological Sciences}
  \textbf{\bibinfo{volume}{362}}, \bibinfo{pages}{933} (\bibinfo{year}{2007}).

\bibitem[{\citenamefont{Mehlhorn et~al.}(2015)\citenamefont{Mehlhorn, Newell,
  Todd, Lee, Morgan, Braithwaite, Hausmann, Fiedler, and
  Gonzalez}}]{mehlhorn15}
\bibinfo{author}{\bibfnamefont{K.}~\bibnamefont{Mehlhorn}},
  \bibinfo{author}{\bibfnamefont{B.~R.} \bibnamefont{Newell}},
  \bibinfo{author}{\bibfnamefont{P.~M.} \bibnamefont{Todd}},
  \bibinfo{author}{\bibfnamefont{M.~D.} \bibnamefont{Lee}},
  \bibinfo{author}{\bibfnamefont{K.}~\bibnamefont{Morgan}},
  \bibinfo{author}{\bibfnamefont{V.~A.} \bibnamefont{Braithwaite}},
  \bibinfo{author}{\bibfnamefont{D.}~\bibnamefont{Hausmann}},
  \bibinfo{author}{\bibfnamefont{K.}~\bibnamefont{Fiedler}}, \bibnamefont{and}
  \bibinfo{author}{\bibfnamefont{C.}~\bibnamefont{Gonzalez}},
  \bibinfo{journal}{Decision} \textbf{\bibinfo{volume}{2}},
  \bibinfo{pages}{191} (\bibinfo{year}{2015}).

\bibitem[{\citenamefont{Simpson et~al.}(2004)\citenamefont{Simpson, Sibly, Lee,
  Behmer, and Raubenheimer}}]{simpson04}
\bibinfo{author}{\bibfnamefont{S.~J.} \bibnamefont{Simpson}},
  \bibinfo{author}{\bibfnamefont{R.~M.} \bibnamefont{Sibly}},
  \bibinfo{author}{\bibfnamefont{K.~P.} \bibnamefont{Lee}},
  \bibinfo{author}{\bibfnamefont{S.~T.} \bibnamefont{Behmer}},
  \bibnamefont{and}
  \bibinfo{author}{\bibfnamefont{D.}~\bibnamefont{Raubenheimer}},
  \bibinfo{journal}{Animal behaviour} \textbf{\bibinfo{volume}{68}},
  \bibinfo{pages}{1299} (\bibinfo{year}{2004}).

\bibitem[{\citenamefont{Hills et~al.}(2012)\citenamefont{Hills, Jones, and
  Todd}}]{hills12}
\bibinfo{author}{\bibfnamefont{T.~T.} \bibnamefont{Hills}},
  \bibinfo{author}{\bibfnamefont{M.~N.} \bibnamefont{Jones}}, \bibnamefont{and}
  \bibinfo{author}{\bibfnamefont{P.~M.} \bibnamefont{Todd}},
  \bibinfo{journal}{Psychological review} \textbf{\bibinfo{volume}{119}},
  \bibinfo{pages}{431} (\bibinfo{year}{2012}).

\bibitem[{\citenamefont{Charnov and Orians}(2006)}]{charnov06}
\bibinfo{author}{\bibfnamefont{E.}~\bibnamefont{Charnov}} \bibnamefont{and}
  \bibinfo{author}{\bibfnamefont{G.~H.} \bibnamefont{Orians}},
  \emph{\bibinfo{title}{Optimal foraging: some theoretical explorations}}
  (\bibinfo{year}{2006}).

\bibitem[{\citenamefont{Bartumeus and Catalan}(2009)}]{bartumeus09}
\bibinfo{author}{\bibfnamefont{F.}~\bibnamefont{Bartumeus}} \bibnamefont{and}
  \bibinfo{author}{\bibfnamefont{J.}~\bibnamefont{Catalan}},
  \bibinfo{journal}{Journal of Physics A: Mathematical and Theoretical}
  \textbf{\bibinfo{volume}{42}}, \bibinfo{pages}{434002}
  (\bibinfo{year}{2009}).

\bibitem[{\citenamefont{B{\'e}nichou et~al.}(2011)\citenamefont{B{\'e}nichou,
  Loverdo, Moreau, and Voituriez}}]{benichou11}
\bibinfo{author}{\bibfnamefont{O.}~\bibnamefont{B{\'e}nichou}},
  \bibinfo{author}{\bibfnamefont{C.}~\bibnamefont{Loverdo}},
  \bibinfo{author}{\bibfnamefont{M.}~\bibnamefont{Moreau}}, \bibnamefont{and}
  \bibinfo{author}{\bibfnamefont{R.}~\bibnamefont{Voituriez}},
  \bibinfo{journal}{Reviews of Modern Physics} \textbf{\bibinfo{volume}{83}},
  \bibinfo{pages}{81} (\bibinfo{year}{2011}).

\bibitem[{\citenamefont{Andrews and Bray}(2004)}]{andrews04}
\bibinfo{author}{\bibfnamefont{S.~S.} \bibnamefont{Andrews}} \bibnamefont{and}
  \bibinfo{author}{\bibfnamefont{D.}~\bibnamefont{Bray}},
  \bibinfo{journal}{Physical biology} \textbf{\bibinfo{volume}{1}},
  \bibinfo{pages}{137} (\bibinfo{year}{2004}).

\bibitem[{\citenamefont{B{\'e}nichou et~al.}(2005)\citenamefont{B{\'e}nichou,
  Coppey, Moreau, Suet, and Voituriez}}]{benichou05}
\bibinfo{author}{\bibfnamefont{O.}~\bibnamefont{B{\'e}nichou}},
  \bibinfo{author}{\bibfnamefont{M.}~\bibnamefont{Coppey}},
  \bibinfo{author}{\bibfnamefont{M.}~\bibnamefont{Moreau}},
  \bibinfo{author}{\bibfnamefont{P.}~\bibnamefont{Suet}}, \bibnamefont{and}
  \bibinfo{author}{\bibfnamefont{R.}~\bibnamefont{Voituriez}},
  \bibinfo{journal}{Physical review letters} \textbf{\bibinfo{volume}{94}},
  \bibinfo{pages}{198101} (\bibinfo{year}{2005}).

\bibitem[{\citenamefont{Newby and Bressloff}(2010)}]{newby10}
\bibinfo{author}{\bibfnamefont{J.~M.} \bibnamefont{Newby}} \bibnamefont{and}
  \bibinfo{author}{\bibfnamefont{P.~C.} \bibnamefont{Bressloff}},
  \bibinfo{journal}{Bulletin of mathematical biology}
  \textbf{\bibinfo{volume}{72}}, \bibinfo{pages}{1840} (\bibinfo{year}{2010}).

\bibitem[{\citenamefont{Viswanathan et~al.}(1996)\citenamefont{Viswanathan,
  Afanasyev, Buldyrev, Murphy et~al.}}]{viswanathan96}
\bibinfo{author}{\bibfnamefont{G.~M.} \bibnamefont{Viswanathan}},
  \bibinfo{author}{\bibfnamefont{V.}~\bibnamefont{Afanasyev}},
  \bibinfo{author}{\bibfnamefont{S.}~\bibnamefont{Buldyrev}},
  \bibinfo{author}{\bibfnamefont{E.}~\bibnamefont{Murphy}},
  \bibnamefont{et~al.}, \bibinfo{journal}{Nature}
  \textbf{\bibinfo{volume}{381}}, \bibinfo{pages}{413} (\bibinfo{year}{1996}).

\bibitem[{\citenamefont{Thornton and Gilden}(2007)}]{thornton07}
\bibinfo{author}{\bibfnamefont{T.~L.} \bibnamefont{Thornton}} \bibnamefont{and}
  \bibinfo{author}{\bibfnamefont{D.~L.} \bibnamefont{Gilden}},
  \bibinfo{journal}{Psychological review} \textbf{\bibinfo{volume}{114}},
  \bibinfo{pages}{71} (\bibinfo{year}{2007}).

\bibitem[{\citenamefont{Posner and Cohen}(1984)}]{posner84}
\bibinfo{author}{\bibfnamefont{M.~I.} \bibnamefont{Posner}} \bibnamefont{and}
  \bibinfo{author}{\bibfnamefont{Y.}~\bibnamefont{Cohen}},
  \bibinfo{journal}{Attention and performance X: Control of language processes}
  \textbf{\bibinfo{volume}{32}}, \bibinfo{pages}{531} (\bibinfo{year}{1984}).

\bibitem[{\citenamefont{Klein}(1988)}]{klein88}
\bibinfo{author}{\bibfnamefont{R.}~\bibnamefont{Klein}},
  \bibinfo{journal}{Nature} \textbf{\bibinfo{volume}{334}},
  \bibinfo{pages}{430} (\bibinfo{year}{1988}).

\bibitem[{\citenamefont{Horowitz and Wolfe}(1998)}]{horowitz98}
\bibinfo{author}{\bibfnamefont{T.~S.} \bibnamefont{Horowitz}} \bibnamefont{and}
  \bibinfo{author}{\bibfnamefont{J.~M.} \bibnamefont{Wolfe}},
  \bibinfo{journal}{Nature} \textbf{\bibinfo{volume}{394}},
  \bibinfo{pages}{575} (\bibinfo{year}{1998}).

\bibitem[{\citenamefont{Wang and Klein}(2010)}]{wang10}
\bibinfo{author}{\bibfnamefont{Z.}~\bibnamefont{Wang}} \bibnamefont{and}
  \bibinfo{author}{\bibfnamefont{R.~M.} \bibnamefont{Klein}},
  \bibinfo{journal}{Vision research} \textbf{\bibinfo{volume}{50}},
  \bibinfo{pages}{220} (\bibinfo{year}{2010}).

\bibitem[{\citenamefont{Smith and Henderson}(2011)}]{smith11}
\bibinfo{author}{\bibfnamefont{T.~J.} \bibnamefont{Smith}} \bibnamefont{and}
  \bibinfo{author}{\bibfnamefont{J.~M.} \bibnamefont{Henderson}},
  \bibinfo{journal}{Journal of Vision} \textbf{\bibinfo{volume}{11}},
  \bibinfo{pages}{3} (\bibinfo{year}{2011}).

\bibitem[{\citenamefont{Bays and Husain}(2012)}]{bays12}
\bibinfo{author}{\bibfnamefont{P.~M.} \bibnamefont{Bays}} \bibnamefont{and}
  \bibinfo{author}{\bibfnamefont{M.}~\bibnamefont{Husain}},
  \bibinfo{journal}{Journal of vision} \textbf{\bibinfo{volume}{12}},
  \bibinfo{pages}{8} (\bibinfo{year}{2012}).

\bibitem[{\citenamefont{Mayer et~al.}(2004)\citenamefont{Mayer, Seidenberg,
  Dorflinger, and Rao}}]{mayer04}
\bibinfo{author}{\bibfnamefont{A.~R.} \bibnamefont{Mayer}},
  \bibinfo{author}{\bibfnamefont{M.}~\bibnamefont{Seidenberg}},
  \bibinfo{author}{\bibfnamefont{J.~M.} \bibnamefont{Dorflinger}},
  \bibnamefont{and} \bibinfo{author}{\bibfnamefont{S.~M.} \bibnamefont{Rao}},
  \bibinfo{journal}{Journal of Cognitive Neuroscience}
  \textbf{\bibinfo{volume}{16}}, \bibinfo{pages}{1262} (\bibinfo{year}{2004}).

\bibitem[{\citenamefont{Horwitz and Newsome}(1999)}]{horwitz99}
\bibinfo{author}{\bibfnamefont{G.~D.} \bibnamefont{Horwitz}} \bibnamefont{and}
  \bibinfo{author}{\bibfnamefont{W.~T.} \bibnamefont{Newsome}},
  \bibinfo{journal}{Science} \textbf{\bibinfo{volume}{284}},
  \bibinfo{pages}{1158} (\bibinfo{year}{1999}).

\bibitem[{\citenamefont{Sapir et~al.}(1999)\citenamefont{Sapir, Soroker,
  Berger, and Henik}}]{sapir99}
\bibinfo{author}{\bibfnamefont{A.}~\bibnamefont{Sapir}},
  \bibinfo{author}{\bibfnamefont{N.}~\bibnamefont{Soroker}},
  \bibinfo{author}{\bibfnamefont{A.}~\bibnamefont{Berger}}, \bibnamefont{and}
  \bibinfo{author}{\bibfnamefont{A.}~\bibnamefont{Henik}},
  \bibinfo{journal}{Nature neuroscience} \textbf{\bibinfo{volume}{2}},
  \bibinfo{pages}{1053} (\bibinfo{year}{1999}).

\bibitem[{\citenamefont{Lepsien and Pollmann}(2002)}]{lepsien02}
\bibinfo{author}{\bibfnamefont{J.}~\bibnamefont{Lepsien}} \bibnamefont{and}
  \bibinfo{author}{\bibfnamefont{S.}~\bibnamefont{Pollmann}},
  \bibinfo{journal}{Journal of cognitive neuroscience}
  \textbf{\bibinfo{volume}{14}}, \bibinfo{pages}{127} (\bibinfo{year}{2002}).

\bibitem[{\citenamefont{Ro et~al.}(2003)\citenamefont{Ro, Farn{\`e}, and
  Chang}}]{ro03}
\bibinfo{author}{\bibfnamefont{T.}~\bibnamefont{Ro}},
  \bibinfo{author}{\bibfnamefont{A.}~\bibnamefont{Farn{\`e}}},
  \bibnamefont{and} \bibinfo{author}{\bibfnamefont{E.}~\bibnamefont{Chang}},
  \bibinfo{journal}{Experimental Brain Research}
  \textbf{\bibinfo{volume}{150}}, \bibinfo{pages}{290} (\bibinfo{year}{2003}).

\bibitem[{\citenamefont{Curtis and D'Esposito}(2003)}]{curtis03}
\bibinfo{author}{\bibfnamefont{C.~E.} \bibnamefont{Curtis}} \bibnamefont{and}
  \bibinfo{author}{\bibfnamefont{M.}~\bibnamefont{D'Esposito}},
  \bibinfo{journal}{Trends in cognitive sciences} \textbf{\bibinfo{volume}{7}},
  \bibinfo{pages}{415} (\bibinfo{year}{2003}).

\bibitem[{\citenamefont{Oh and Kim}(2004)}]{oh04}
\bibinfo{author}{\bibfnamefont{S.-H.} \bibnamefont{Oh}} \bibnamefont{and}
  \bibinfo{author}{\bibfnamefont{M.-S.} \bibnamefont{Kim}},
  \bibinfo{journal}{Psychonomic bulletin \& review}
  \textbf{\bibinfo{volume}{11}}, \bibinfo{pages}{275} (\bibinfo{year}{2004}).

\bibitem[{\citenamefont{Melville}(1999)}]{melville99}
\bibinfo{author}{\bibfnamefont{H.}~\bibnamefont{Melville}},
  \emph{\bibinfo{title}{Bartleby, the scrivener}} (\bibinfo{publisher}{State
  University of NY-Albany}, \bibinfo{year}{1999}).

\bibitem[{gar()}]{gary17}
\urlprefix\url{http://www.garystpc.com/pest-id/gophers-moles.php}.

\bibitem[{\citenamefont{Handford}(1987)}]{handford87}
\bibinfo{author}{\bibfnamefont{M.}~\bibnamefont{Handford}},
  \emph{\bibinfo{title}{Where's Waldo?}} (\bibinfo{publisher}{Little, Brown
  Boston}, \bibinfo{year}{1987}).

\bibitem[{eco()}]{ecot}
\urlprefix\url{http://www.cs.colorado.edu/department/maps/ec.html}.

\bibitem[{\citenamefont{Aksay et~al.}(2001)\citenamefont{Aksay, Gamkrelidze,
  Seung, Baker, and Tank}}]{aksay01}
\bibinfo{author}{\bibfnamefont{E.}~\bibnamefont{Aksay}},
  \bibinfo{author}{\bibfnamefont{G.}~\bibnamefont{Gamkrelidze}},
  \bibinfo{author}{\bibfnamefont{H.}~\bibnamefont{Seung}},
  \bibinfo{author}{\bibfnamefont{R.}~\bibnamefont{Baker}}, \bibnamefont{and}
  \bibinfo{author}{\bibfnamefont{D.}~\bibnamefont{Tank}},
  \bibinfo{journal}{Nature neuroscience} \textbf{\bibinfo{volume}{4}},
  \bibinfo{pages}{184} (\bibinfo{year}{2001}).

\bibitem[{\citenamefont{McNaughton et~al.}(2006)\citenamefont{McNaughton,
  Battaglia, Jensen, Moser, and Moser}}]{mcnaughton06}
\bibinfo{author}{\bibfnamefont{B.~L.} \bibnamefont{McNaughton}},
  \bibinfo{author}{\bibfnamefont{F.~P.} \bibnamefont{Battaglia}},
  \bibinfo{author}{\bibfnamefont{O.}~\bibnamefont{Jensen}},
  \bibinfo{author}{\bibfnamefont{E.~I.} \bibnamefont{Moser}}, \bibnamefont{and}
  \bibinfo{author}{\bibfnamefont{M.-B.} \bibnamefont{Moser}},
  \bibinfo{journal}{Nature Reviews Neuroscience} \textbf{\bibinfo{volume}{7}},
  \bibinfo{pages}{663} (\bibinfo{year}{2006}).

\bibitem[{\citenamefont{Burak and Fiete}(2009)}]{burak09}
\bibinfo{author}{\bibfnamefont{Y.}~\bibnamefont{Burak}} \bibnamefont{and}
  \bibinfo{author}{\bibfnamefont{I.~R.} \bibnamefont{Fiete}},
  \bibinfo{journal}{PLoS Comput Biol} \textbf{\bibinfo{volume}{5}},
  \bibinfo{pages}{e1000291} (\bibinfo{year}{2009}).

\bibitem[{\citenamefont{Engbert et~al.}(2011)\citenamefont{Engbert,
  Mergenthaler, Sinn, and Pikovsky}}]{engbert11}
\bibinfo{author}{\bibfnamefont{R.}~\bibnamefont{Engbert}},
  \bibinfo{author}{\bibfnamefont{K.}~\bibnamefont{Mergenthaler}},
  \bibinfo{author}{\bibfnamefont{P.}~\bibnamefont{Sinn}}, \bibnamefont{and}
  \bibinfo{author}{\bibfnamefont{A.}~\bibnamefont{Pikovsky}},
  \bibinfo{journal}{Proceedings of the National Academy of Sciences}
  \textbf{\bibinfo{volume}{108}}, \bibinfo{pages}{E765} (\bibinfo{year}{2011}).

\bibitem[{\citenamefont{Amari}(1977)}]{amari77}
\bibinfo{author}{\bibfnamefont{S.}~\bibnamefont{Amari}},
  \bibinfo{journal}{Biol. Cybern.} \textbf{\bibinfo{volume}{27}},
  \bibinfo{pages}{77} (\bibinfo{year}{1977}), ISSN \bibinfo{issn}{0340-1200
  (Print)}.

\bibitem[{\citenamefont{Seung}(1996)}]{seung96}
\bibinfo{author}{\bibfnamefont{H.~S.} \bibnamefont{Seung}},
  \bibinfo{journal}{Proceedings of the National Academy of Sciences}
  \textbf{\bibinfo{volume}{93}}, \bibinfo{pages}{13339} (\bibinfo{year}{1996}).

\bibitem[{\citenamefont{Zhang}(1996)}]{zhang96}
\bibinfo{author}{\bibfnamefont{K.}~\bibnamefont{Zhang}}, \bibinfo{journal}{The
  journal of neuroscience} \textbf{\bibinfo{volume}{16}}, \bibinfo{pages}{2112}
  (\bibinfo{year}{1996}).

\bibitem[{\citenamefont{Coombes and Owen}(2005)}]{coombes05b}
\bibinfo{author}{\bibfnamefont{S.}~\bibnamefont{Coombes}} \bibnamefont{and}
  \bibinfo{author}{\bibfnamefont{M.}~\bibnamefont{Owen}},
  \bibinfo{journal}{Physical Review Letters} \textbf{\bibinfo{volume}{94}},
  \bibinfo{pages}{148102} (\bibinfo{year}{2005}).

\bibitem[{\citenamefont{Kilpatrick and Bressloff}(2010)}]{kilpatrick10c}
\bibinfo{author}{\bibfnamefont{Z.~P.} \bibnamefont{Kilpatrick}}
  \bibnamefont{and} \bibinfo{author}{\bibfnamefont{P.~C.}
  \bibnamefont{Bressloff}}, \bibinfo{journal}{Physica D: Nonlinear Phenomena}
  \textbf{\bibinfo{volume}{239}}, \bibinfo{pages}{1048} (\bibinfo{year}{2010}).

\bibitem[{\citenamefont{Bressloff}(2012)}]{bressloff12}
\bibinfo{author}{\bibfnamefont{P.~C.} \bibnamefont{Bressloff}},
  \bibinfo{journal}{Journal of Physics A: Mathematical and Theoretical}
  \textbf{\bibinfo{volume}{45}}, \bibinfo{pages}{033001}
  (\bibinfo{year}{2012}).

\bibitem[{\citenamefont{Coombes and Schmidt}(2010)}]{coombes10}
\bibinfo{author}{\bibfnamefont{S.}~\bibnamefont{Coombes}} \bibnamefont{and}
  \bibinfo{author}{\bibfnamefont{H.}~\bibnamefont{Schmidt}},
  \bibinfo{journal}{Discrete and Continuous Dynamical Systems. Series S}
  (\bibinfo{year}{2010}).

\bibitem[{\citenamefont{Horowitz and Wolfe}(2003)}]{horowitz03}
\bibinfo{author}{\bibfnamefont{T.}~\bibnamefont{Horowitz}} \bibnamefont{and}
  \bibinfo{author}{\bibfnamefont{J.}~\bibnamefont{Wolfe}},
  \bibinfo{journal}{Visual Cognition} \textbf{\bibinfo{volume}{10}},
  \bibinfo{pages}{257} (\bibinfo{year}{2003}).

\bibitem[{\citenamefont{Stemmler et~al.}(1995)\citenamefont{Stemmler, Usher,
  and Niebur}}]{stemmler95}
\bibinfo{author}{\bibfnamefont{M.}~\bibnamefont{Stemmler}},
  \bibinfo{author}{\bibfnamefont{M.}~\bibnamefont{Usher}}, \bibnamefont{and}
  \bibinfo{author}{\bibfnamefont{E.}~\bibnamefont{Niebur}},
  \bibinfo{journal}{Science} \textbf{\bibinfo{volume}{269}},
  \bibinfo{pages}{1877} (\bibinfo{year}{1995}).

\bibitem[{\citenamefont{Angelucci and Bressloff}(2006)}]{angelucci06}
\bibinfo{author}{\bibfnamefont{A.}~\bibnamefont{Angelucci}} \bibnamefont{and}
  \bibinfo{author}{\bibfnamefont{P.~C.} \bibnamefont{Bressloff}},
  \bibinfo{journal}{Progress in brain research} \textbf{\bibinfo{volume}{154}},
  \bibinfo{pages}{93} (\bibinfo{year}{2006}).

\bibitem[{\citenamefont{Bressloff}(2001)}]{bressloff01}
\bibinfo{author}{\bibfnamefont{P.~C.} \bibnamefont{Bressloff}},
  \bibinfo{journal}{Physica D: Nonlinear Phenomena}
  \textbf{\bibinfo{volume}{155}}, \bibinfo{pages}{83} (\bibinfo{year}{2001}).

\bibitem[{\citenamefont{Coombes and Laing}(2011)}]{coombes11}
\bibinfo{author}{\bibfnamefont{S.}~\bibnamefont{Coombes}} \bibnamefont{and}
  \bibinfo{author}{\bibfnamefont{C.}~\bibnamefont{Laing}},
  \bibinfo{journal}{Physical Review E} \textbf{\bibinfo{volume}{83}},
  \bibinfo{pages}{011912} (\bibinfo{year}{2011}).

\bibitem[{\citenamefont{Avitabile and Schmidt}(2015)}]{avitabile15}
\bibinfo{author}{\bibfnamefont{D.}~\bibnamefont{Avitabile}} \bibnamefont{and}
  \bibinfo{author}{\bibfnamefont{H.}~\bibnamefont{Schmidt}},
  \bibinfo{journal}{Physica D: Nonlinear Phenomena}
  \textbf{\bibinfo{volume}{294}}, \bibinfo{pages}{24} (\bibinfo{year}{2015}).

\bibitem[{\citenamefont{Silvester}(2000)}]{silvester00}
\bibinfo{author}{\bibfnamefont{J.~R.} \bibnamefont{Silvester}},
  \bibinfo{journal}{The Mathematical Gazette} \textbf{\bibinfo{volume}{84}},
  \bibinfo{pages}{460} (\bibinfo{year}{2000}).

\bibitem[{\citenamefont{Ermentrout}(1998)}]{ermentrout98}
\bibinfo{author}{\bibfnamefont{B.}~\bibnamefont{Ermentrout}},
  \bibinfo{journal}{Reports on progress in physics}
  \textbf{\bibinfo{volume}{61}}, \bibinfo{pages}{353} (\bibinfo{year}{1998}).

\bibitem[{\citenamefont{Coombes et~al.}(2012)\citenamefont{Coombes, Schmidt,
  and Bojak}}]{coombes12}
\bibinfo{author}{\bibfnamefont{S.}~\bibnamefont{Coombes}},
  \bibinfo{author}{\bibfnamefont{H.}~\bibnamefont{Schmidt}}, \bibnamefont{and}
  \bibinfo{author}{\bibfnamefont{I.}~\bibnamefont{Bojak}}, \bibinfo{journal}{J
  Math Neurosci} \textbf{\bibinfo{volume}{2}}, \bibinfo{pages}{9}
  (\bibinfo{year}{2012}).

\bibitem[{\citenamefont{Kilpatrick and Ermentrout}(2013)}]{kilpatrick13}
\bibinfo{author}{\bibfnamefont{Z.~P.} \bibnamefont{Kilpatrick}}
  \bibnamefont{and}
  \bibinfo{author}{\bibfnamefont{B.}~\bibnamefont{Ermentrout}},
  \bibinfo{journal}{SIAM J. Appl. Dyn. Syst.} \textbf{\bibinfo{volume}{12}},
  \bibinfo{pages}{61} (\bibinfo{year}{2013}).

\bibitem[{\citenamefont{Olton et~al.}(1977)\citenamefont{Olton, Collison, and
  Werz}}]{olton77}
\bibinfo{author}{\bibfnamefont{D.~S.} \bibnamefont{Olton}},
  \bibinfo{author}{\bibfnamefont{C.}~\bibnamefont{Collison}}, \bibnamefont{and}
  \bibinfo{author}{\bibfnamefont{M.~A.} \bibnamefont{Werz}},
  \bibinfo{journal}{Learning and Motivation} \textbf{\bibinfo{volume}{8}},
  \bibinfo{pages}{289} (\bibinfo{year}{1977}).

\bibitem[{\citenamefont{Floresco et~al.}(1997)\citenamefont{Floresco, Seamans,
  and Phillips}}]{floresco97}
\bibinfo{author}{\bibfnamefont{S.~B.} \bibnamefont{Floresco}},
  \bibinfo{author}{\bibfnamefont{J.~K.} \bibnamefont{Seamans}},
  \bibnamefont{and} \bibinfo{author}{\bibfnamefont{A.~G.}
  \bibnamefont{Phillips}}, \bibinfo{journal}{Journal of Neuroscience}
  \textbf{\bibinfo{volume}{17}}, \bibinfo{pages}{1880} (\bibinfo{year}{1997}).

\bibitem[{\citenamefont{Poll and Kilpatrick}(2016)}]{poll16}
\bibinfo{author}{\bibfnamefont{D.~B.} \bibnamefont{Poll}} \bibnamefont{and}
  \bibinfo{author}{\bibfnamefont{Z.~P.} \bibnamefont{Kilpatrick}},
  \bibinfo{journal}{Journal of Statistical Mechanics: Theory and Experiment}
  \textbf{\bibinfo{volume}{2016}}, \bibinfo{pages}{053201}
  (\bibinfo{year}{2016}).

\bibitem[{\citenamefont{Laing and Troy}(2003)}]{laing03}
\bibinfo{author}{\bibfnamefont{C.~R.} \bibnamefont{Laing}} \bibnamefont{and}
  \bibinfo{author}{\bibfnamefont{W.~C.} \bibnamefont{Troy}},
  \bibinfo{journal}{SIAM Journal on Applied Dynamical Systems}
  \textbf{\bibinfo{volume}{2}}, \bibinfo{pages}{487} (\bibinfo{year}{2003}).

\bibitem[{\citenamefont{Folias and Bressloff}(2004)}]{folias04}
\bibinfo{author}{\bibfnamefont{S.~E.} \bibnamefont{Folias}} \bibnamefont{and}
  \bibinfo{author}{\bibfnamefont{P.~C.} \bibnamefont{Bressloff}},
  \bibinfo{journal}{SIAM Journal on Applied Dynamical Systems}
  \textbf{\bibinfo{volume}{3}}, \bibinfo{pages}{378} (\bibinfo{year}{2004}).

\bibitem[{\citenamefont{Owen et~al.}(2007)\citenamefont{Owen, Laing, and
  Coombes}}]{owen07}
\bibinfo{author}{\bibfnamefont{M.}~\bibnamefont{Owen}},
  \bibinfo{author}{\bibfnamefont{C.}~\bibnamefont{Laing}}, \bibnamefont{and}
  \bibinfo{author}{\bibfnamefont{S.}~\bibnamefont{Coombes}},
  \bibinfo{journal}{New Journal of Physics} \textbf{\bibinfo{volume}{9}},
  \bibinfo{pages}{378} (\bibinfo{year}{2007}).

\bibitem[{\citenamefont{G{\"o}k{\c{c}}e
  et~al.}(2017)\citenamefont{G{\"o}k{\c{c}}e, Avitabile, and
  Coombes}}]{gokcce17}
\bibinfo{author}{\bibfnamefont{A.}~\bibnamefont{G{\"o}k{\c{c}}e}},
  \bibinfo{author}{\bibfnamefont{D.}~\bibnamefont{Avitabile}},
  \bibnamefont{and} \bibinfo{author}{\bibfnamefont{S.}~\bibnamefont{Coombes}},
  \bibinfo{journal}{Journal of Mathematical Neuroscience}
  (\bibinfo{year}{2017}).

\bibitem[{\citenamefont{Poll and Kilpatrick}(2015)}]{poll15}
\bibinfo{author}{\bibfnamefont{D.}~\bibnamefont{Poll}} \bibnamefont{and}
  \bibinfo{author}{\bibfnamefont{Z.~P.} \bibnamefont{Kilpatrick}},
  \bibinfo{journal}{SIAM Journal on Applied Mathematics}
  \textbf{\bibinfo{volume}{75}}, \bibinfo{pages}{1553} (\bibinfo{year}{2015}).

\bibitem[{\citenamefont{Deneve et~al.}(1999)\citenamefont{Deneve, Latham, and
  Pouget}}]{deneve99}
\bibinfo{author}{\bibfnamefont{S.}~\bibnamefont{Deneve}},
  \bibinfo{author}{\bibfnamefont{P.~E.} \bibnamefont{Latham}},
  \bibnamefont{and} \bibinfo{author}{\bibfnamefont{A.}~\bibnamefont{Pouget}},
  \bibinfo{journal}{Nature Neuroscience} \textbf{\bibinfo{volume}{2}},
  \bibinfo{pages}{741} (\bibinfo{year}{1999}).

\bibitem[{\citenamefont{Olsen et~al.}(2010)\citenamefont{Olsen, Bhandawat, and
  Wilson}}]{olsen10}
\bibinfo{author}{\bibfnamefont{S.~R.} \bibnamefont{Olsen}},
  \bibinfo{author}{\bibfnamefont{V.}~\bibnamefont{Bhandawat}},
  \bibnamefont{and} \bibinfo{author}{\bibfnamefont{R.~I.}
  \bibnamefont{Wilson}}, \bibinfo{journal}{Neuron}
  \textbf{\bibinfo{volume}{66}}, \bibinfo{pages}{287} (\bibinfo{year}{2010}).

\bibitem[{\citenamefont{Beck et~al.}(2011)\citenamefont{Beck, Latham, and
  Pouget}}]{beck11}
\bibinfo{author}{\bibfnamefont{J.~M.} \bibnamefont{Beck}},
  \bibinfo{author}{\bibfnamefont{P.~E.} \bibnamefont{Latham}},
  \bibnamefont{and} \bibinfo{author}{\bibfnamefont{A.}~\bibnamefont{Pouget}},
  \bibinfo{journal}{Journal of Neuroscience} \textbf{\bibinfo{volume}{31}},
  \bibinfo{pages}{15310} (\bibinfo{year}{2011}).

\bibitem[{\citenamefont{Churchland et~al.}(2012)\citenamefont{Churchland,
  Cunningham, Kaufman, Foster, Nuyujukian, Ryu, and Shenoy}}]{churchland12}
\bibinfo{author}{\bibfnamefont{M.~M.} \bibnamefont{Churchland}},
  \bibinfo{author}{\bibfnamefont{J.~P.} \bibnamefont{Cunningham}},
  \bibinfo{author}{\bibfnamefont{M.~T.} \bibnamefont{Kaufman}},
  \bibinfo{author}{\bibfnamefont{J.~D.} \bibnamefont{Foster}},
  \bibinfo{author}{\bibfnamefont{P.}~\bibnamefont{Nuyujukian}},
  \bibinfo{author}{\bibfnamefont{S.~I.} \bibnamefont{Ryu}}, \bibnamefont{and}
  \bibinfo{author}{\bibfnamefont{K.~V.} \bibnamefont{Shenoy}},
  \bibinfo{journal}{Nature} \textbf{\bibinfo{volume}{487}}, \bibinfo{pages}{51}
  (\bibinfo{year}{2012}).

\end{thebibliography}

\end{document}